\documentclass[journal]{vgtc}                     % final (journal style)
\onlineid{0}

\vgtccategory{Research}

\vgtcpapertype{please specify}

\title{\vispur: Visual Aids for Identifying and Interpreting Spurious Associations in Data-Driven Decisions}

\author{%
  \authororcid{Xian Teng}{0000-0003-2289-2882},
  \authororcid{Yongsu Ahn}{0000-0002-5797-5445}, and 
  \authororcid{Yu-Ru Lin}{0000-0002-8497-3015}
}

\authorfooter{
  \item
  	Xian Teng is with University of Pittsburgh.
  	E-mail: xit22@pitt.edu.
  \item
  	Yongsu Ahn is with University of Pittsburgh.
  	E-mail: yongsu.ahn@pitt.edu.

  \item Yu-Ru Lin is with University of Pittsburgh.
  	E-mail: yurulin@pitt.edu.
}

  \abstract{{{Big data and machine learning tools have jointly empowered humans in making data-driven decisions. However, many of them capture empirical associations that might be spurious due to confounding factors and subgroup heterogeneity.}
  {The famous Simpson's paradox is such a phenomenon where aggregated and subgroup-level associations contradict with each other, causing cognitive confusions and difficulty in making adequate interpretations and decisions.}
  {Existing tools provide little insights for humans to locate, reason about, and prevent pitfalls of spurious association in practice.}
  {We propose \vispur, a visual analytic system that provides a causal analysis framework and a human-centric workflow for tackling spurious associations.}
  {These include a \confounderdashboard, which can automatically identify possible confounding factors, and a \subgroupviewer, which allows for the visualization and comparison of diverse subgroup patterns that likely or potentially result in a misinterpretation of causality. Additionally, we propose a \reasoningstoryboard, which uses a flow-based approach to illustrate paradoxical phenomena, as well as an interactive \decisiondiagnosis panel that helps ensure accountable decision-making.}
  {Through an expert interview and a controlled user experiment, our qualitative and quantitative results demonstrate that the proposed ``de-paradox'' workflow and the designed visual analytic system are effective in helping human users to identify and understand spurious associations, as well as to make accountable causal decisions.}
}}


\teaser{
  \centering
\includegraphics[width=\textwidth]{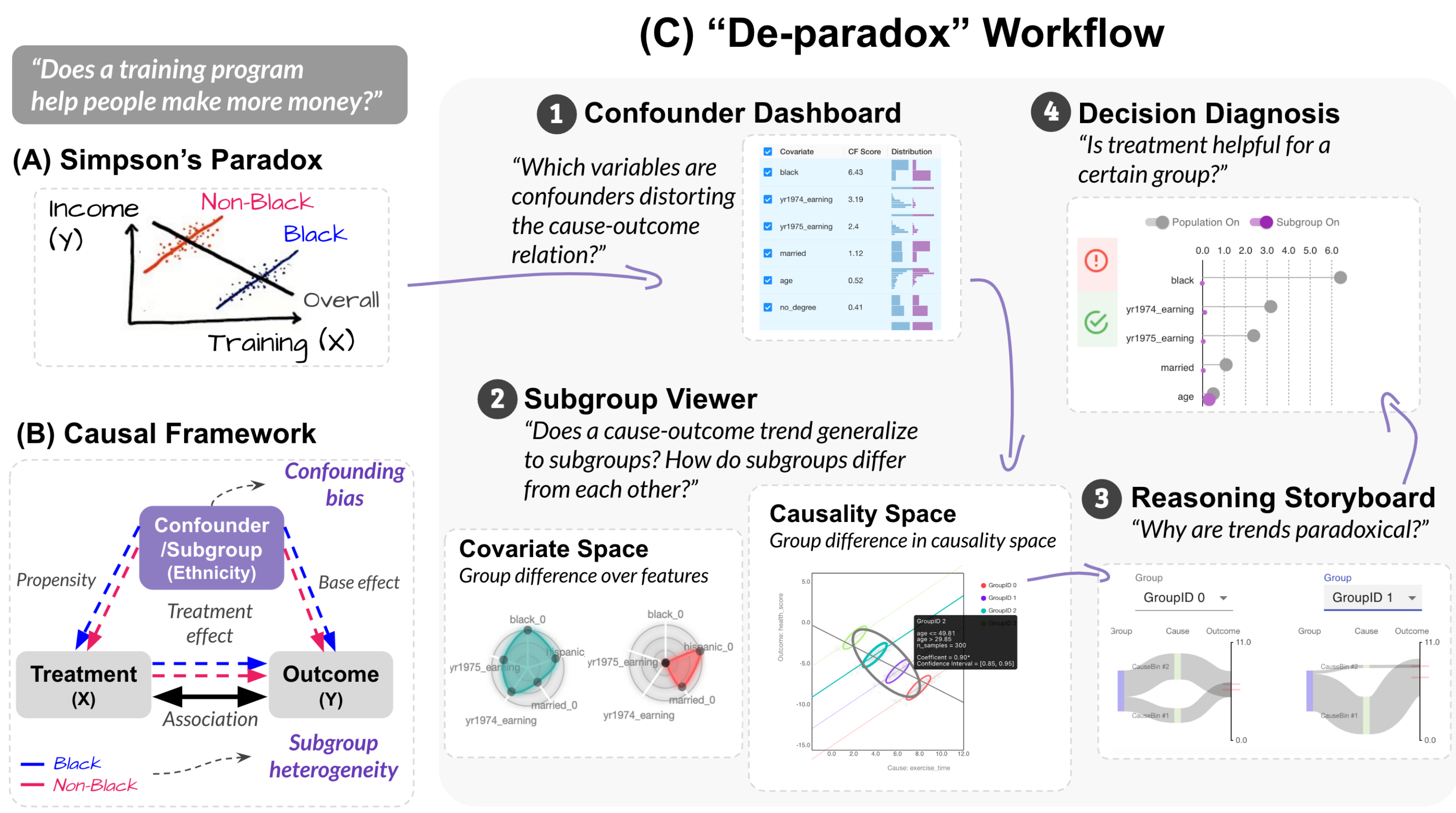}
  \caption{{This work provides a ``de-paradox'' workflow to help analyze observational data and overcome spurious and paradoxical associations that can lead to misleading interpretations of causal effects.}
  {(A) Spurious associations, including Simpson's paradox, are prevalent in observational studies. E.g., in a study that investigates the effect of a job training program, the cause (training program) and outcome (earnings) can be distorted by a third variable (ethnicity), leading to a misleading interpretation of the causal effect. (B) We identify two major sources for spuriousness: (1) confounding bias and (2) subgroup heterogeneity, based on causal literature. 
  {(C) We propose a novel ``de-paradox'' workflow to tackle the two major sources of spuriousness, which include (C.1) \confounderdashboard, (C.2) \subgroupviewer, (C.3) \reasoningstoryboard, and (C.4) \decisiondiagnosis, to guide users to ``de-paradox'' and reason about the most appropriate causal interpretation of association paradox.}
  }}

  \label{fig:teaser}
}

\renewcommand{\manuscriptnotetxt}{}

\newcommand{\tocomments}[1]{{\color{black}{#1}}}

\graphicspath{{figs/}{figures/}{pictures/}{images/}{./}} % where to search for the images

\usepackage{tabu}                      % only used for the table example
\usepackage{booktabs}                  % only used for the table example
\usepackage{lipsum}                    % used to generate placeholder text
\usepackage{mwe}                       % used to generate placeholder figures
\usepackage{amsmath}                   % used to generate equations
\usepackage{float}                     % figure position specifier
\usepackage{enumitem} \setlist{noitemsep}
\usepackage{adforn}
\usepackage{graphicx}
\usepackage{graphics}
\usepackage{mathptmx}                  % use matching math font

\usepackage{xspace}
\usepackage{soul}
\usepackage[svgnames]{xcolor}

\definecolor{lightgreen}{HTML}{7fc97f}
\definecolor{lightpurple}{HTML}{984ea3}
\definecolor{lightorange}{HTML}{fdc086}
\definecolor{lightblue}{HTML}{386cb0}

\newcommand{\conebox}[1]{\colorbox{lightpurple}{#1}}
\newcommand{\ctwobox}[1]{\colorbox{lightgreen}{#1}}
\newcommand{\cthreebox}[1]{\colorbox{lightorange}{#1}}
\newcommand{\cfourbox}[1]{\colorbox{lightblue}{#1}}
\newcommand{\cone}[1]{\textcolor{lightpurple}{#1}}
\newcommand{\ctwo}[1]{\textcolor{lightgreen}{#1}}
\newcommand{\cthree}[1]{\textcolor{lightorange}{#1}}
\newcommand{\cfour}[1]{\textcolor{lightblue}{#1}}

\newcommand{\vispur}{{\scshape \color{black} Vispur}\xspace}

\newcommand{\confounderdashboard}{{\scshape \color{black} Confounder Dashboard}\xspace}
\newcommand{\partition}{{\scshape \color{black} Subgroup Partition}\xspace}
\newcommand{\subgroupviewer}{{\scshape \color{black} Subgroup Viewer}\xspace}
\newcommand{\decisiondiagnosis}{{\scshape \color{black} Decision Diagnosis}\xspace}
\newcommand{\reasoningstoryboard}{{\scshape \color{black} Reasoning Storyboard}\xspace}

\newcommand{\covariatespace}{{\bf Covariate Space}\xspace}
\newcommand{\causalityspace}{{\bf Causality Space}\xspace}
\newcommand{\basicstatistics}{{\bf Basic Statistics}\xspace}
\newcommand{\imbalancechart}{{\bf Imbalance Chart}\xspace}

\begin{document}

%%%%%%%%%%%%%%%%%%%%%%%%%%%%%%%%%%%%%%%%%%%%%%%%%%%%%%%%%%%%%%%%
%%%%%%%%%%%%%%%%%%%%%% START OF THE PAPER %%%%%%%%%%%%%%%%%%%%%%
%%%%%%%%%%%%%%%%%%%%%%%%%%%%%%%%%%%%%%%%%%%%%%%%%%%%%%%%%%%%%%%%

% \firstsection{Introduction}\label{sec:introduction}

\maketitle

\section{Introduction}\label{sec:introduction}
{Decision-making processes in a variety of domains concern estimates of causal impacts of a shift in policy via what-if question, such as changes in product pricing for business or new treatments for health professionals \cite{cookson2021equity,kievit2013simpson}.}
{Despite the ample capability of big data and machine learning tools available for data-driven decisions, many data analysis practices and ML methods often pick up on {\it spurious associations} (referred as ``shortcuts'' in ML models) instead of learning the true {\it causal relationships} \cite{degrave2021ai,joshi2022all,park2021comparison}.}
{The lack of causal insights in empirical association analysis can cause confusion and difficulty in decision-making process \cite{lerman2018computational,kievit2013simpson}, and even pose a risk on societal efficacy \cite{jeffrey2021israeli}, equity \cite{obermeyer2019dissecting,park2021comparison}, and fairness \cite{lerman2018computational,cookson2021equity}.}
% an overall association might mask heterogeneous patterns among subgroups \cite{von2021simpson}, and paradoxical/conflicting associations (e.g., Simpson's paradox) might cause confusion and difficulty in decision-making \cite{lerman2018computational,kievit2013simpson}.}
% it might reflect social disparity in accessing health resources \cite{park2021comparison},
% {The lack of causal insights in association analysis would raise societal and public health risk regarding {efficacy} \cite{erica2021nearly,jeffrey2021israeli}, {equity} \cite{obermeyer2019dissecting,park2021comparison} and {fairness} \cite{verma2022impacts,cookson2021equity}.}

% about the Lalonde data
% The National Supported Work (NSW) Demonstration was a federally-funded program implemented in the mid-1970s, with the objective of providing work experience for a period of 12 to 18 months to individuals who had faced economic and social problems prior to enrollment in the program.
% The National Supported Work Demon- stration (NSW) was a temporary employment program designed to help disad- vantaged workers lacking basic job skills move into the labor market by giving them work experience and counseling in a sheltered environment.

{\bf Motivating Example.} To illustrate this issue, consider an example of an educational dataset\footnote{The Lalonde dataset (1986) \cite{lalonde1986evaluating,dehejia2002propensity,dehejia1999causal} is from the National Supported Work Demonstration, concerning a government-funded job training program that aimed at help citizens to increase their earnings. More information can be seen at \url{https://users.nber.org/~rdehejia/nswdata2.html}.}, illustrated in Fig.~\ref{fig:teaser}A, which concerns whether a government-funded training program is effective in helping individuals earn more money  \cite{lalonde1986evaluating,dehejia2002propensity,dehejia1999causal}. 
{In the data, program participants earn less money than non-participants, which may raise doubts about the program's effectiveness and lead to its termination. However, a closer look reveals Simpson's paradox when considering ethnicity. When looking at Black and non-Black subgroups separately, program participants actually earn more than non-participants. This paradox arises probably because Black participants face greater disadvantages, leading to lower earnings but a higher participation rate driven by their need for financial improvement. Combining the subgroups creates an overall negative trend due to participants' lower earnings. This example highlights the importance of investigating the spurious program-earning association and avoiding hasty program termination that could further disadvantage already marginalized groups.}

{Like this example, many spurious (possibly paradoxical) associations are hard for humans to analyze and reason about. In recent years, causal machine learning has made progress in estimate individualized and subgroup-level causal responses (e.g., Black/non-Black) by incorporating ML models into causal framework \cite{athey2019generalized,kunzel2019metalearners,oprescu2019orthogonal,syrgkanis2019machine}
% \cite{wager2018estimation,athey2019generalized,kunzel2019metalearners,mackey2018orthogonal,oprescu2019orthogonal,syrgkanis2019machine}
{However, these methods tend to have strong structural assumptions imposed on data generation process, which poses a challenge for decision-makers and data practitioners to choose the best model as well as to make a valid judgement on the estimated results.} 
{Besides, these models encapsulate causal computations into a black box without an explanation on how covariates influence treatment or how treatment/covariates affect outcome, thus providing little insights on  subgroups that may have conflicting associations.}
{Recent visual analytic systems have been developed to promote interpretability for causal analysis on multidimensional data \cite{jin2020visual,xie2020visual,xie2020causalflow}. But they neither detect spurious associations, nor explain -- for a target causal relation -- what are the causal reasons behind paradoxical associations among subgroups.}}
% neither rigorously inspect or explain spuriousness under a causal framework, nor reveal varied subgroup behaviors in the space of causal space.}}

% {To tackle this problem, there have been two lines of works: (1) visualization interfaces for causal or/and subgroup analysis, (2) causal machine learning.}
% {The first category of works develop visual interfaces to let users explore causal relationships, clusters, or model behaviors over subsets of data. But they neither specify a formal causal model to locate and explain spurious associations, nor investigate subgroup characteristics in the space of causality.} 
% {The second category is method-oriented. They incorporate ML models into causal framework to estimate individualized (or conditional) causal responses from data. Such algorithms encapsulates causal computations into a black box, making it for humans to fully understand and trust the results.}
% {Overall, existing works provide little insight to identify a spurious association (depends on humans to locate through domain knowledge), reveal/interpret the causal root of its emergence, and they do not investigate subgroup characteristics in the scope of causality.}

{% a summary of task
Given the profound impact of spurious association and the  research gap, we propose a systematic causal analysis of spurious associations on the basis of a formal causal model \cite{imbens2010rubin} by developing an interactive visual system.}
{% a high-level summary of our work
By allowing users to see contrastive patterns -- features or causality-related behaviors -- between treatment arms and among different subgroups of data, the workflow guides practitioners to detect, reason about causal sources of a spurious association, and to overcome common pitfalls in data-driven decision-making.}
{% Specifically, the causal diagram depicts ...
Specifically, we focus on two causal mechanisms: {\bf confounding bias} and {\bf subgroup heterogeneity}, as they widely exist in observational studies \cite{lerman2018computational,kievit2013simpson} and have significant decision implications (e.g., the job training program example). As shown in Fig.~\ref{fig:teaser}B, confounding bias indicates the distortion effect from confounding variables (e.g., ethnicity) that might simultaneously affect cause and outcome. Subgroup heterogeneity refers to subgroup patterns in the space of causality including the {\it propensity} towards treatment \cite{rosenbaum1983central}, {\it base effect} towards outcome, along with subgroup-level {\it causal effects}.}
To transform causal theory into practical use, we interview three domain experts, including an educational system designer, a social worker, and a financial data scientist to identify a set of design requirements that existing tools/practices fail to support. We further propose a ``de-paradox'' workflow with four major components shown in Fig.~\ref{fig:teaser}C. 
The workflow allows users to identify possible confounding factors (Fig.~\ref{fig:teaser}C.1), compare subgroup patterns (Fig.~\ref{fig:teaser}C.2), hypothesize and reason about paradoxical phenomena (Fig.~\ref{fig:teaser}C.3), and perform responsible decision-making such as whether to impose a treatment or not (Fig.~\ref{fig:teaser}C.4).

%These include (C.1) a \confounderdashboard, which can automatically identify possible confounding factors involved in a cause-outcome relationship, and (C.2) a \subgroupviewer, which allows for the visualization and comparison of diverse subgroup patterns in both attribute and causality space (e.g., propensity, base effects, etc). Additionally, we have designed (C.3) a flow-based \reasoningstoryboard, by illustrating the pathways from treatments to outcome to allow human users to hypothesize and reason about paradoxical phenomena, as well as (C.4) an interactive \decisiondiagnosis panel that helps ensure responsible decision-making such as whether to impose a treatment or not.}

{% VISPUR
To facilitate such a workflow, we develop \vispur\footnote{The code is available at: \url{https://github.com/picsolab/VISPUR}}, \textbf{\underline{vis}}ualizing s\textbf{\underline{pur}}ious associations, a visual analytic system to enable causal analysis of spurious associations. The system incorporates a suite of statistical techniques, algorithms, and visual components to help identify causal roots of spurious associations, as well as modules to reason about association reversal/paradox and to make informed decisions. To summarize, our contributions include:
}
\begin{itemize}
    \item {{\bf A systematic workflow that incorporates the design needs of our target users to help them navigate the causal analysis of spurious associations.}} \tocomments{Our target users are data practitioners or domain experts who need to answer a causal question but lack causal inference knowledge.} We close the gap between causal theory and practical use by identifying a set of design guidelines and proposing a systematic workflow. \tocomments{Our work explores the visual analytic design issues concerning causal analysis and interpretation of spurious associations from empirical data.}
    \item {{\bf \vispur, a visual analytic system that investigates the causal sources of spurious associations or paradoxes by utilizing visualizations that reduce human's cognitive burdens.} {We present two visual views: \subgroupviewer produces glyphs (``visual signatures'') that encode multidimensional data features; It also incorporates a causal space where key causal concepts are simultaneously revealed and compared; \reasoningstoryboard communicates causal stories through event pathways to support humans users in the process of paradox reasoning.}}
    \item {{\bf Evaluations to demonstrate the utility of our system.}} We conduct a controlled user study and an expert interview study, showing that \vispur not only enables users to better locate causal roots of a spurious association (confounders and causal behaviors among subsets of data), but also allows them to better understand why a paradoxical association emerges whilst aggregating subgroups. These observations from \vispur together lead to a richer understanding of the data.
\end{itemize}

% {The remainder of this paper is organized as follows. Section~\ref{sec:relatedwork} provides a review of related works. Section~\ref{sec:designguideline} highlights design challenges and requirements. Section~\ref{sec:methodology} describes our causal framework, metrics, and algorithms. The \vispur system design is discussed in Section~\ref{sec:systemdesign}. We report evaluation results in Section~\ref{sec:userstudy} and give an extended discussion in Section ~\ref{sec:discussion}. Finally, we conclude this work in Section~\ref{sec:conclusion}.}

\section{Related Work}\label{sec:relatedwork}
% {We first describe the causal framework our work is based upon, and then review two types of related visual systems, including visual causal analysis, and visual subgroup analysis.} 
\subsection{Causal Inference Framework}
% \tocomments{Although randomized controlled trials (RCTs) are considered the ``gold standard'' to establish causality, they are often unethical, impractical, or untimely \cite{guo2021vaine,guo2023causalvis}. Causal inference based on observational data has been widely applied in health domain \cite{cookson2021equity}, social, and political sciences \cite{clark2015big,gerring2005causation}. The randomization design of RCTs ensures that subjects from two treatment groups have comparable characteristics, namely, covariates are {\it balanced}. In contrast, in observational studies two treatment groups might have very distinct feature distributions, and the outcome difference might eventually trace back to factors other than treatment, making it questionable to endow association with a causal interpretation.}

\tocomments{Randomized controlled trials (RCTs) are the ``gold standard'' for causality, but they are often unethical, impractical, or untimely \cite{guo2021vaine,guo2023causalvis}. In the absence of RCTs, causal inference based on observational data has been extensively utilized in many domains \cite{cookson2021equity, clark2015big, gerring2005causation}. RCTs ensure comparable characteristics, {\it balanced}, between treatment groups through randomization, while observational studies may have distinct feature distributions ({\it imbalanced}), making it possible that outcome difference might eventually trace back to factors other than treatment.}

\tocomments{The potential outcomes framework, also called the Rubin Causal Model (RCM) \cite{imbens2010rubin}, is a theoretical framework for causal inference in both observational and experimental studies. Consider again the example in Introduction~\ref{sec:introduction}, it involves defining two potential outcomes for each person -- the potential outcome had they participated in the program and the outcome had they not. By comparing outcome differences across subjects, the average treatment effect (ATE) is estimated. Since it is impossible to observe {\it both} potential outcomes (as one of the potential outcomes is always missing in reality), additional assumptions are necessary for inferring the treatment effect. These assumptions in our study include overlap (or positivity) \cite{heckman1997matching}, which assumes that participants could have chosen not to attend the program and vice versa, and the stable unit treatment value assumption (SUTVA) \cite{vanderweele2013causal,cole2009consistency}, which assumes that subjects do not influence each other's participation decisions and employment outcomes, and there are no hidden variations of treatment that might lead to distinct outcomes. Additionally, the unconfoundedness assumption \cite{rosenbaum1983central} requires that all confounders should have been measured for causal analysis.}

\tocomments{Our system is designed in the context of RCM framework on the basis of above assumptions, allowing users to investigate spurious or paradoxical association in observational studies.}
\subsection{Visual Causal Analysis}

% {Researchers have been developing visual analytic tools to support causal reasoning to overcome the lack of decision support in practice with raw data and statistical results \toRone{alone} \cite{chen2011data}. Existing works focus on several aspects, including (a) human causality perception from general-purpose visualization \cite{yen2019exploratory, xiong2019illusion, kale2021causal, kadaba2007visualizing}, (b) design of novel visual representations \cite{baker2001good,rum1980magic,armstrong2014visualizing, rucker2008simpson,friendly2013elliptical}, (b) visual analytic systems for exploratory analysis of complex causal relations \cite{wang2015visual,wang2017visual,xie2020visual,jin2020visual,dang2015reactionflow,wongsuphasawat2012exploring,lu2017visual,xie2020causalflow}.}

Researchers have been developing visual analytic tools to support causal reasoning to overcome the lack of decision support in practice with raw data and statistical results \tocomments{alone} \cite{chen2011data}. Existing works focus on various aspects such as (a) human causality perception from general-purpose visualization \cite{yen2019exploratory, kale2021causal}, (b) design of new visual representations \cite{baker2001good,rum1980magic,armstrong2014visualizing, rucker2008simpson,friendly2013elliptical}, (b) visual analytic systems for causal discovery \cite{dang2015reactionflow, wang2015visual, xie2020causalflow, xie2020visual, jin2020visual}, as well as \tocomments{(c) open-source visual tools and libraries for causal inference \cite{greifer2020covariate,shimoni2019evaluation,chen2020causalml,econml,guo2023causalvis,guo2021vaine}.} 

Researchers have found that general-purpose visualizations can lead to error-prone perceived causality due to confirmation bias and information overload \cite{yen2019exploratory}. Different visual encodings, such as bar charts and icon arrays, do not significantly enhance causal inferences beyond contingency tables \cite{kale2021causal}. Given the limitations of general-purpose visualizations, researchers have designed new representations by focusing on Simpson's paradox. These designs include B-K diagram \cite{baker2001good}, platform scale representation \cite{rum1980magic}, comet chart \cite{armstrong2014visualizing}, circle-line plots \cite{rucker2008simpson} and data ellipse diagram \cite{friendly2013elliptical} mainly for pedagogical purpose. Recent advancements in visual interfaces have facilitated exploratory analysis of complex causal relations, specifically in large-scale multidimensional  \cite{wang2015visual, wang2017visual, xie2020visual} and sequence data \cite{dang2015reactionflow,xie2020causalflow,jin2020visual}, incorporating state-of-the-art automatic causality discovery algorithms.
% {For example, Causality Explorer \cite{xie2020visual} applied causal graph detection algorithms on big data, designed an uncertainty-aware node-link causal graph, and provides interactive what-if analysis and simulations of potential actions (e.g., change of variable values). The seqCausal \cite{jin2020visual} designed a flow-based visualization with the Granger causality algorithm \cite{granger1969investigating} to show how event sequences progress among key events in the causal graph.}
% {Causal Structure Investigator (CSI) \cite{wang2017visual} further addressed the subgroup analysis problem where users could explore data partition (either manual partition or clustering algorithms) and infer various causal models. Users could pool all causal models to summarize common causal relations as well as compare to recognize pattern differences.}
{However, those systems lack interpretability and do not locate spuriousness, or reason about the causal roots of a misleading association.}
\tocomments{In recent years, various open-source visual tools and packages have emerged to support causal inference. These include Causalvis \cite{guo2023causalvis}, VAINE \cite{guo2021vaine}, and a Causal AI Suite \cite{kiciman2022causal} integrating DoWhy \cite{sharma2020dowhy}, EconML \cite{econml}, Causica\footnote{Causica: \url{https://github.com/microsoft/causica/}}, and ShowWhy \cite{microsoft2022introduction}. While these tools have similar functionalities with our system, such as confounder estimation, covariate balance checking, and subgroup visualizations, their intended audience and goals differ from ours. They primarily target causal inference experts and support the iterative causal inference process ranging from confounder investigation, matching and weighting, inference and reporting. In contrast, our target users are data analysts and domain experts who might lack causal inference knowledge and experience in utilizing causal inference packages. Our \vispur interface is designed to provide a user-friendly platform for those encountering distortion and paradox without causality background. While ShowWhy \cite{microsoft2022introduction} offers a codeless user interface for a broader audience, it does not focus on explaining Simpson's paradox by revealing causal mechanisms. VAINE \cite{guo2021vaine}, closely related to our system, detects Simpson's paradox through clustering algorithms and provides cause-outcome and covariate views on a cluster level. In contrast, our \vispur system distinguishes itself by its capability of interpreting Simpson's paradox in terms of two causal mechanisms: confounding bias and heterogeneous subgroups. By incorporating causal analysis components, such as confounder identification and diagnosis component, our system enables users to gain a better understanding of the paradox, and prevent them from mistakenly interpreting overlaid regression lines as causal effects.}

\subsection{Visual Analytics for Subgroup Analysis}

Subgroup analysis has been a popular topic in data-driven decisions, because an aggregated pattern is not always generalized to (even differ from) that of subgroups. Visual analytic systems have been designed to support a variety of subgroup-level analysis, such as visualizing clusters and features \cite{kwon2017clustervision,furmanova2020taggle,blumenschein2018smartexplore,ahn2022tribe}, \tocomments{understanding event sequences or disease progressions \cite{jin2020visual,kwon2020dpvis}, as well as examining model behaviors over subsets of data \cite{dingen2018regressionexplorer, kwon2022rmexplorer, wexler2019if, cabrera2019fairvis, gleicher2020boxer, cheng2020dece}.} For example, Taggle \cite{furmanova2020taggle} is a tabular design to visualize high-dimensional data in terms of record clusters, subspaces, correlations, and pattern similarity across different levels of stacked aggregation.
% { In contrast to directly showing records and features in a table, many systems attempt to visualize clusters with various clustering algorithms through different views (e.g., scatter plots, parallel axes, etc).}
% {ClusterVision \cite{kwon2017clustervision} is designed to compare the clustering results obtained from a series of competing clustering techniques, and rank them through five quality metrics encoded in a radar chart.}
\tocomments{DPVis \cite{kwon2020dpvis} focuses on event sequence data and allows users to investigate heterogeneous disease progression pathways of patient subgroups.} \tocomments{In addition to static/sequential data properties, many tools have been developed to understand the performance of machine learning models over subsets of data. Examples include DECE \cite{cheng2020dece},} What-If Tool \cite{wexler2019if}, Boxer \cite{gleicher2020boxer}, and FairVIS \cite{cabrera2019fairvis} among others. Relevant systems include RegressionExplorer \cite{dingen2018regressionexplorer} and RMExplorer \cite{kwon2022rmexplorer}. RegressionExplorer is tailored for logistic regression model analysis, supporting dynamic subgroup generation and visualizations of subgroup-level parameters. RMExplorer enables users to define patient subgroups based on various characteristics and assess the performance and fairness of risk models on these subgroups \cite{kwon2022rmexplorer}. \tocomments{DECE \cite{cheng2020dece} shares design similarities with our system, as it enables users to create subgroups using multi-feature decision rules and provides contrastive feature comparison through side-by-side histograms.} However, there is a limited amount of work on developing visual analytic systems specifically for investigating subgroup patterns in the space of causality.

{Our system \vispur examines heterogeneous subgroup patterns on the basis of a causal framework (Fig.~\ref{fig:teaser}B), revealing how likely different subgroups take a treatment, how likely they obtain a target outcome, along with the subgroup-level cause-outcome relations. Given those information, \vispur explains how subgroup-level behaviors are linked to an overall spurious association or a Simpson's paradox.}

\section{Design Requirements and Tasks}\label{sec:designguideline}

% In this section, we highlight challenges in reasoning about spurious associations and present design guidelines for building an analytic tool.
\tocomments{{\bf Target Users and Interviews.} Our target users are data practitioners, specifically those who are not experts in the field of causality. They may need to explore causal questions or provide empirical evidence to decision makers. For instance, one of our domain experts sought to understand the extent to which their educational system improved students' programming skills. While they may not possess causality knowledge or use causal inference tools, they should have a basic understanding of statistics and apply statistical techniques (such as tests, associations, and regression analysis) in their daily jobs. To ensure our system meets their needs, we interviewed three expert users from diverse backgrounds, including a social worker ({\bf P1}), a trading analyst ({\bf P2}), and an educational system designer ({\bf P3}).}
Each section lasted about an hour in the form of a semi-structured interview. {To understand the interviewees' knowledge/experience of statistics and causality, as well as their obstacles in addressing counterintuitive associations and/or paradox phenomenon, we engaged them to} consider a task where they perform association analysis using empirical data. We encouraged them to apply their current practices/tools, and reflect on the limitations of those methods and expectations for future systems.  More details of interviews are provided in Supplementary Materials~\ref{sec:appendix}.

{\bf Common Challenges.} Expressed by interviewees, they all have encountered counterintuitive or paradoxical associations in different studies. 
\tocomments{For example, the educational system designer {\bf P3} reported an {\it unexpected} association -- the more engaged students have been, the worse their performance were in skill evaluation.}
% For example, {\bf P3} reported an {\it unexpected} association from their education data, showing that the more code samples students have studied, the worse their performance were in skill evaluation.
However, the current common strategies (e.g., statistical tests, regressions using R, Python libraries) fail to detect spurious associations or help understand the reasons why paradoxical/spurious associations emerge. The identified four major challenges are (see Table S1 in Supplementary Materials~\ref{sec:appendix} for more details):
\begin{enumerate}
    \item[\conebox{\bf C1.}] {\bf Unable to \underline{identify} possible covariates that might distort a cause-outcome relationship} (mentioned by {\bf P1, P2}). Interviewees have admitted that their current strategies were arbitrary in choosing the covariates to adjust/control in data analysis (e.g., regression). There is no or little reflection of causal relationships among a rich set of variables in their current workflow.
    \item[\ctwobox{\bf C2.}] {\bf Unable to \underline{detect} subgroups and \underline{examine} their characteristics in face of a rich set of variables} (mentioned by {\bf P1, P2, P3}). 
    Our interviewees attempted to conduct subgroup analysis to reveal data heterogeneity. For instance, {\bf P2} hypothesized that different students might engage with the educational system differently, thereafter their skill could be affected by the system at distinct extents. But no tools could help test this hypothesis. They choose either not to execute subgroup analysis, or to rely on prior experiences for manual partition. Systematic and automatic methods to discover subgroups are still lacking.
    \item[\cthreebox{\bf C3.}] {\bf Unable to \underline{interpret} the changes of associations, the counterintuitive associations, as well as the emergence of a Simpson's paradox, given different data partition or regression scenarios} (mentioned by {\bf P2, P3}). Interviewees acknowledged that ``associations do not necessarily imply causation,'' but they were still bringing causal explanations to make sense of the derived coefficients (and $p-$values). If an association went against their causal expectation, they would feel confused. They often got further confused by inconsistent coefficients obtained from multiple trials of testing different possible regression models with distinct subsets of data.
    \item[\cfourbox{\bf C4.}] \tocomments{\bf Unable to \underline{make} a confident decision because a given association (even obtained from a subset of data) may be distorted and contain mixed subgroups.} (mentioned by {\bf P1-P3}).
    % {\bf Hard to \underline{make} a confident decision, given inconsistent, paradoxical, and counterintuitive associations} (mentioned by {\bf P1, P2}).
    When drawing a conclusion, such as whether or not the designed educational system is beneficial for students, \tocomments{{\bf P3} admitted they cannot confidently make a claim regarding their system's effectiveness, ``{\it it is hard to trust any associations, even after dividing data into subsets.}''}
\end{enumerate}
\tocomments{In summary, {\bf C1-C2} are concerned with the challenges of identifying and discovering confounding bias and heterogeneous subgroups from data, whereas {\bf C3-C4} are concerned with how to interpret and make informed decisions for a given use scenario. The challenges thus result in different design requirements.}

{\bf Design Requirements} \label{sec:designguideline:requirement}
To solve the above challenges ({\bf C}), we identified four major requirements ({\bf R}), with five specific design tasks ({\bf T}).
\begin{enumerate}
    \item[{\bf R1.}] {\bf Facilitate identification of confounders (motivated by \conebox{\bf C1}).}
    {\bf (T1)} As a cause-outcome relationship might be spurious when additional confounders are involved (Fig.~\ref{fig:teaser}B), a visual analytic tool should guide users in reflecting causal relationships among variables, i.e., which covariates might simultaneously affect the preference for treatment and for outcome. It should also provide quantitative measurements and intuitive visualizations for human users to locate the most likely confounders.
    \item[{\bf R2.}] {\bf Support the discovery, characterization and examination of subgroup heterogeneity (motivated by \ctwobox{\bf C2}).}
    
    {\bf (T2)} The system should facilitate both manual (hypothesis-driven) and automatic (data-driven) subgroup discovery. Users should be able to define subgroups manually using one or multiple covariates. It should also incorporate algorithms to automatically search for subgroups that are internally homogeneous but differ from each other.
    
    {\bf (T3)} The system should enable users to examine the heterogeneity of subgroups from a variety of aspects, such as feature properties, to what extent a subgroup is likely to take certain treatment actions ({\it propensity}), to what extent a subgroup is to take certain outcome scenarios ({\it base effect}), what are the subgroup-level {\it associations}, whether they are distorted because of nested confounding bias ({\it causal effect}).
    
    \item[{\bf R3.}] {\bf Enable the interpretation of a spurious association and/or Simpson's paradox (motivated by \cthreebox{\bf C3}).}
    
    {\bf (T4)} The visual analytic system should enable users to understand why the aggregation of subgroups with different characteristics could lead to a paradoxical phenomenon. \tocomments{In particular, the two causal mechanisms -- confounding bias, and subgroup heterogeneity -- should be explained in a clear and intuitive way.}
    
    \item[{\bf R4.}] \tocomments{\bf Facilitate ``spuriousness'' diagnosis for any given association as well as accountable decision-making (motivated by \cfourbox{\bf C4}).}
    
    {\bf (T5)} The system should enable users to make adequate judgements regarding whether a chosen association is spurious or not. It should also provide quantitative information and visual encodings to assist users in making accountable decisions about how the cause might influence the target outcome, overall or given a certain subgroup.
\end{enumerate}
\section{Methods: Causal Framework, Metrics, Algorithms}\label{sec:methodology}
We present the problem setup, metrics of confounding bias and imbalance, as well as a propensity tree algorithm to discover subgroups.

\subsection{Setup}\label{sec:methodology:diagram}
% {Our study builds upon the causal framework that include mechanisms and assumptions about spurious associations and causal inference. First, the analysis of spurious associations is designed to explore its two mechanisms: i) The divergence of association against causal effect. In theory, a causal effect depicts how outcome may differ under treatment conditions (e.g., treated vs control), where any other ``risk'' factors of outcome are {comparable} between treatment arms. However, in observational studies, this prerequisite is almost never satisfied when studying associations. Taking an aforementioned motivating example as shown in Fig.~\ref{fig:teaser}B, ethnicity is a confounder that not only affects a subject's choice of participation ({\it propensity}), but also plays as a ``risk'' factor that influences a subject's annual income ({\it base effect}). The distortion due to confounders is referred as {\it confounder bias}, where the outcome difference might trace back to ethnicity difference instead of to treatment difference. ii) The mixed effect of {subgroups}. As shown in Fig.~\ref{fig:teaser}B, subgroups might have various patterns in the space of causality, reflected by the three causal arrows ({\it propensity}, {\it base effect}, {\it causal effect}) and an observable cause-outcome association. An overall pattern does not generalize to subgroups.}

\tocomments{Following the potential outcome framework \cite{imbens2010rubin}, we postulate the existence of a pair of potential outcomes for each subject $(Y_i(1), Y_i(0))$, with the individual causal effect defined as $\tau_i=Y_i(1)-Y_i(0)$ and the ATE defined as $\tau = \mathrm{mean}_{i}[Y_i(1)-Y_i(0)]$. Let $X_i$ be the binary indicator for the treatment, with $X_i=0$ indicating that subject $i$ received the control treatment and $X_i=1$ indicating that $i$ received the active treatment. The realized outcome for subject $i$ is the potential outcome corresponding to the treatment received: $Y_i = X_iY_i(1)+(1-X_i)Y_i(0)$. The pair of $(Y_i(1), Y_i(0))$ could never be observed at the same time. Let $\mathbf{Z}_i$ to be the $K$-dimensional vector of covariates, or pretreatment variables, known not to be affected by the treatment.}

\tocomments{For the purpose of illustration, we plot an ``adapted'' directed acyclic graph (DAG) \cite{pearl1995causal} in Fig.~\ref{fig:teaser}B, where vertices represent variables $(X,Y,\mathbf{Z})$, dashed directed arrows represent causal relations, and solid double-ended arrows represent associations. The causal arrow $\mathbf{Z}\rightarrow X$ represents {\bf treatment propensity} \cite{rosenbaum1983central}, indicating the inclination of subjects or a subset defined by $\mathbf{Z}$ to receive a specific treatment. The causal arrow $\mathbf{Z}\rightarrow Y$ represents the {\bf base effect}, capturing subjects' inherent tendency to have a particular outcome without receiving active treatment. The arrow $X\rightarrow Y$ represents the true {\bf causal effect}. The double-ended arrow $X\leftrightarrow Y$ represents the {\bf cause-outcome association}.}

\tocomments{In Fig.~\ref{fig:teaser}B, {\bf confounding bias} means that $\mathbf{Z} \not\!\perp\!\!\!\perp X$ since $\mathbf{Z}$ affects $X$, hence the feature distributions, $\mathbf{Z}|X=0$ and $\mathbf{Z}|X=1$, are not the same ({\bf imbalanced}). Therefore, the outcome difference of two treatment arms might trace back to difference in $\mathbf{Z}$ instead of to $X$. The other mechanism {\bf heterogeneous subgroups} suggests that subgroups hold very distinct {propensity}, {base effect}, {causal effect}, as well as {cause-outcome associations} in the space of causality (e.g., two sets of arrows colored as red and blue). An overall pattern does not generalize to subgroups. }

\tocomments{The problem setup generalizes to a continuous treatment. The potential outcome framework assumes the existence of a set of potential outcomes $Y_i(x), x\in\Omega$, where $\Omega$ denotes the set of all possible values of treatment \cite{austin2019assessing}. Typically, a ``dose-response'' function will be estimated $Y=g_{\theta}(X)$ where $\theta$ depicts causal effect. We fit a regression model to estimate $g_{\theta}$ based on outcome data type, e.g., a linear model for continuous outcomes and a logistic model for binary outcomes.}

\subsection{Measuring Confounding Bias}\label{sec:methodology:cf_metrics}
% In order to facilitate the identification of confounding variable $Z$, we formulate a quantitative metric called {\bf confounding tendency (CT) score}. Basically, CF compares to what extent the cause-outcome relationship would change before and after controlling for $Z$ \cite{norton2015simpson}. Without loss of generality, we focus on a continuous cause $X$ and a binary outcome $Y$. We fit two logistic regression models: $\mathrm{logit}[p(Y=1)] = \beta_0 + \beta_1 X$, versus, $\mathrm{logit}[p(Y=1)] = \beta_0^\prime + \beta_1^\prime X + \beta_z Z$. Then we compute the change of odds ratio estimate as $\mathrm{CF}({Z}) = |e^{\beta_1^\prime} - e^{\beta_1}| / e^{\beta_1}$ to indicator how strong $Z$ confounds the cause-outcome relationship.

In order to facilitate the identification of a potential \tocomments{confounding variable $Z$ (one dimension from $\mathbf{Z}$), we follow \cite{norton2015simpson} quantitatively compute the extent to which the cause-outcome relationship would change before and after controlling for $Z$, and refer it as {\bf \underline{c}on\underline{f}ounding score} (CF score)}. \tocomments{Suppose the outcome $Y$ is binary, we then fit two logistic regression models: (a) one with just the treatment as a predictor $\mathrm{logit}[p(Y=1)] = \beta_0 + \beta_1 X$, and (b) the other with the potential confounder included as a covariate $\mathrm{logit}[p(Y=1)] = \beta_0^\prime + \beta_1^\prime X + \beta_z Z$.} Then we compute the change of odds ratio \tocomments{for the treatment between the two models} $\mathrm{CF}({Z}) = |e^{\beta_1^\prime} - e^{\beta_1}| / e^{\beta_1}$ to indicate how strong $Z$ confounds the cause-outcome relationship.

\subsection{Measuring Covariate Imbalance}\label{sec:methodology:imbalance}
% The notion of {\bf covariate (im)balance} is proposed to capture how comparable the treated/untreated groups are in terms of other pretreatment variables \cite{belitser2011measuring,austin2015moving,austin2009balance,austin2019assessing}. A high imbalance score given by $Z$ (e.g., age) suggests that the treated/untreated subjects hold very different age distributions, therefore their differences in outcome might be largely explained by the gap in $Z$.

\tocomments{To measure to what extent the $\mathbf{Z}$ are imbalanced, existing works have proposed a range of metrics to measure {\bf covariate (im)balance} \cite{austin2015moving,austin2019assessing}.} A high imbalance score given by $Z$ (e.g., age) suggests that the treated/untreated subjects hold very different age distributions, therefore their differences in outcome might be largely explained by the gap in $Z$. \tocomments{When $X$ is dichotomous, existing works \cite{imai2014covariate,fong2018covariate} have shown that an ideal covariate balance could be operationalized as $\mathrm{E}[f(\mathbf{Z})|X=1]=\mathrm{E}[f(\mathbf{Z})|X=0]$, where $f$ is an arbitrary vector-valued measurable function whose expectation exists. When $X$ is continuous, the ideal balance can be operationalized as $\mathrm{E}[\mathbf{Z}]\mathrm{E}[X]=0$ \cite{fong2018covariate}. For practical computation, when $X$ is binary we use the standardized mean difference proposed in \cite{austin2015moving} to measure imbalance:} $d = {(\bar{Z}_1 - \bar{Z}_0}) / {\sqrt{0.5(s_1^2 + s_0^2)}}$, where $\bar{Z}_1, \bar{Z}_0$ are covariate means for $X=1$ and $X=0$, $s_1^2, s_0^2$ are their variances accordingly. When $X$ is continuous, we use correlation-based metrics proposed in \cite{fong2018covariate}, $d = \mathrm{Spearman}(Z,X)$, to measure the per-feature imbalance.

\subsection{Algorithm for Subgroup Discovery}\label{sec:methodology:algorithm}
Our visual analytic system incorporates ML methods to automatically discover subgroups to be shown on the visualization interfaces. Several existing works have used tree-based methods for searching and estimating subgroup-level treatment effects \cite{athey2016recursive,wager2018estimation}. These methods start by recursively splitting the feature space until they have partitioned data into a set of leaves (subgroups) $\mathcal{L}$, each of which, $l \in \mathcal{L}$, contains a subset of data points. One might consider that the data points belonging to the same leaf, $i \in l$, act as if they had come from a randomized experiment. Then a leaf-specific effect size $\tau(l)$ is computed by comparing the difference of outcomes $\tau(l) = \bar{Y}_{l,1}-\bar{Y}_{l,0}$, or learning a dose-response relation $Y_{l}=g_{\theta}(X_{l})$. \tocomments{Suppose $g_{\theta}$ takes a logistic regression form $g_{\theta}(X_{l}) = 1/(1+e^{-(\beta_0(l) + \beta_1(l)X_{l})})$, the coefficient $\beta_1(l)$ could be used to represent the effect size $\tau(l)$ for the leaf $l$.} \vispur exploits the state-of-the-art subgroup partition algorithm: propensity tree \cite{wager2018estimation,athey2016recursive}. It combines decision tree techniques with propensity scores in causal inference. It partitions data based on features $\mathbf{Z}$ while using $X$ as the target to ensure a balanced distribution of treated and control units within each subgroup. This approach helps address confounding and enables reliable estimation of causal effects within specific subgroups. It is different from decision tree in a way that an estimated effect size is attached to each of the leaves rather than predicted values of treatment. We adopt the propensity tree algorithm because: (i) It is particularly useful in observational studies, where we want to minimize confounding bias due to variance in treatment propensity; (ii) It is able to handle a large size of features by automatically selecting the most ``important'' features to split on.

\section{\vispur: System Components} \label{sec:systemdesign}

\begin{figure*}[ht]
    \centering
    \includegraphics[width=0.85\linewidth]{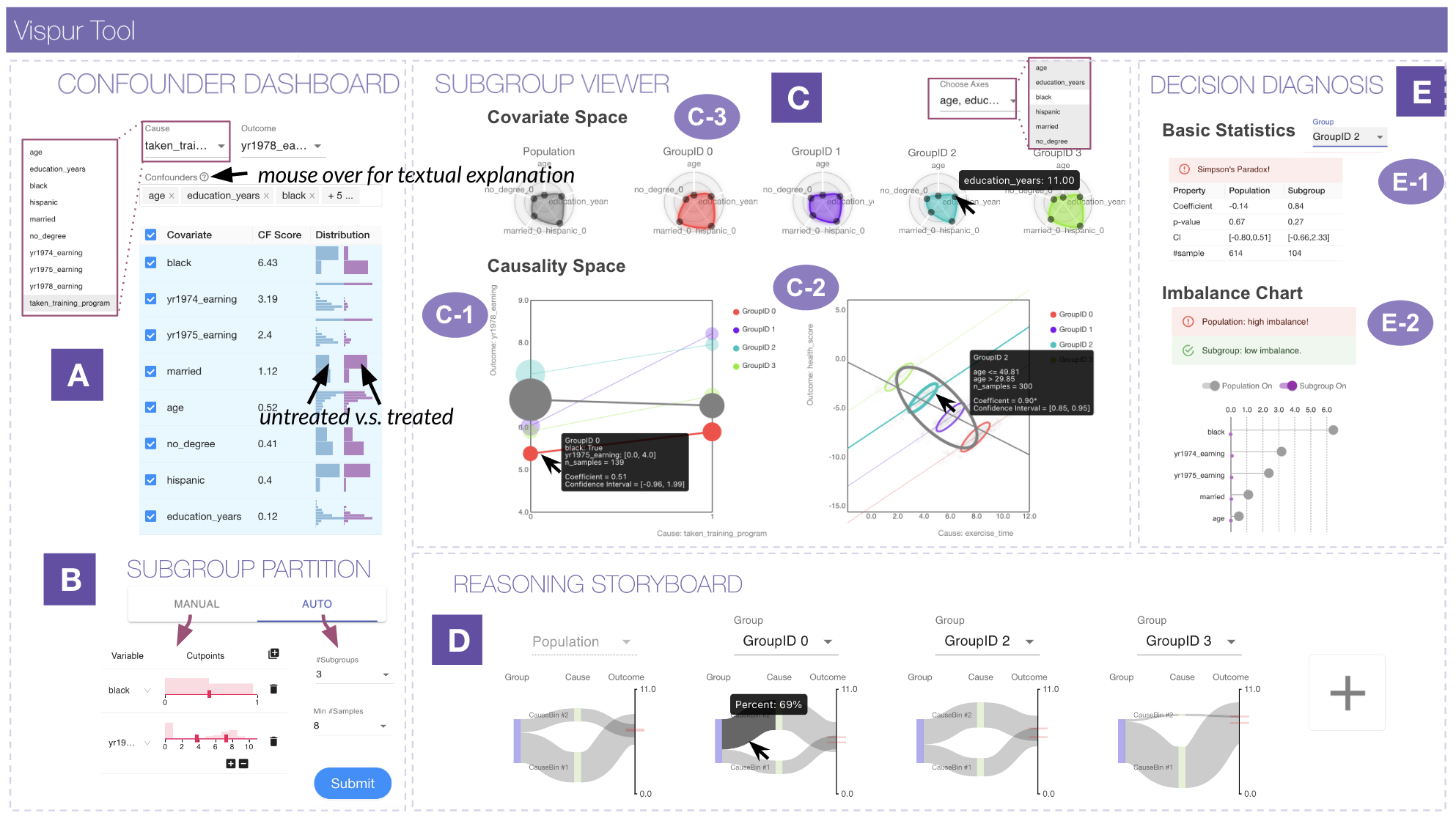}
    \caption{The \vispur system includes several components: (A) the \confounderdashboard assists with selecting confounders by comparing CF scores and distributions across treatment arms, (B) the \partition supports manual or automated subgroup generation by setting partition rules or algorithm parameters, (C) the \subgroupviewer includes two separate spaces---{\causalityspace} \tocomments{(C-1 for binary treatment, C-2 for continuous treatment)} and {\covariatespace} (C-3)---for displaying subgroups' causal/associational properties and selected features, (D) the \reasoningstoryboard allows for subgroup selection and comparison through story flows, and (E) the \decisiondiagnosis provides detailed subgroup-level statistics (E-1), warning messages about confounding bias, and association distortion (E-2).}
    \label{fig:system_overview}
\end{figure*}

% {An overview of \vispur system. (A) The {\bf \confounderdashboard} allows selecting cause and outcome, ranking remaining covariates and displaying the comparison of distributions across treatment arms to assist users in selecting confounders. (B) The {\bf \partition} supports both manual and automated subgroup generation by either allowing users to specify partition rules (manual) or algorithm parameters (auto). (C) The {\bf \subgroupviewer} contains two separate spaces: {\causalityspace} (C-1 for binary treatment, C-2 for continuous treatment) and {\covariatespace} (C-3), where the former demonstrates subgroups' causal/assocaitional properties such as {\it propensity}, {\it base effect}, as well as cause-outcome {\it association} for generated subgroups, while the latter space displays a set of radar-shaped glyphs for subgroups which encode the median values of selected features. (D) The {\bf \reasoningstoryboard} allows users to select and compare subgroups of interest by investigating story flows from treatment actions to outcome status. (E) The {\bf \decisiondiagnosis} reports the detailed subgroup-level statistics in {\basicstatistics} (E-1) and {\imbalancechart} (E-2) for a selected subgroup. It pops out warning messages to alert users about the presence of confounding bias and association distortion.}

{In light of the design requirements, we describe how we design \vispur to implement the aforementioned ``de-paradox'' workflow for causal analysis of spurious/paradoxical associations. \tocomments{The demonstration is based on the Lalonde dataset \cite{lalonde1986evaluating}, as it has been a benchmark data in causal inference literature and incorporated for use cases in many libraries such as MatchIt \cite{stuart2011matchit} and Cobalt \cite{greifer2020covariate}.} As shown in Fig.~\ref{fig:system_overview}A-E, our visual design leverages human perceptional features to reduce users' cognitive burdens, highlighting two major components---\subgroupviewer and \reasoningstoryboard---to assist inspecting subgroup characteristics, as well as making sense of association paradox, respectively. (1) In \subgroupviewer, we exploit radar-shaped glyphs to encode rich and multidimensional data features into a small and compact graphic representation, allowing users to visually compare the most important ``feature signatures'' of subgroups (Fig.~\ref{fig:system_overview}C-3). To distinguish subgroups' causal properties, we redefine the most popular two-dimensional space (that data practitioners are familiar with) to represent a new causality space (Fig.~\ref{fig:system_overview}C-1, C-2), where several causal concepts (propensity, base effects, etc) are encoded into simple visual signals, facilitating a straightforward and contrastive comparison. Furthermore, to support paradox reasoning, we leverage flow-based visualizations based on a common idea that the cause is followed by the effect. The progression pathways showcase how subjects chose their preferred treatment options, and how alternative chains of treatments lead to different outcomes (Fig.~\ref{fig:system_overview}D).}

% Users will explore the {\bf \generator} (A) to identify confounding factors that might distort a cause-outcome relationship ({\bf R1 -- confounder identification}), then they will make use of {\bf \partition} (B) to produce a set of meaningful subgroups to further examine their differences {\bf \group}, from the aspects of both causality and features ({\bf R2} -- subgroup heterogeneity). In face of possible paradoxical or conflicting associations, users will interact with {\bf \storyboard} (D) for detailed explanations ({\bf R3} -- reasoning). Finally, users will refer to the interface {\bf \diagnosis} (E) when making a decision regarding the effect of a treatment ({\bf R4} -- decision-making). That said, our system assists users answer a series of questions in the workflow of studying empirical associations. The distinguishing contribution of our work lies in its ability -- visually and interactively -- to locate and explain association spuriousness ({\bf \storyboard}), as well as to explicitly reveal the heterogeneous subgroup patterns from both aspects of causality and features ({\bf \partition} and {\bf \group}).
\subsection{\confounderdashboard}\label{sec:systemdesign:confounderdashboard}
The \confounderdashboard interface in Fig.~\ref{fig:system_overview}A enables users to set up cause/outcome variables and to locate the most likely confounders that are distorting the specified cause-outcome relationship ({\bf T1}). The flexibility of the system allows users to tailor their investigations to their specific research interests. For example, if someone is curious about the effectiveness of the job training programs in boosting annual earnings, they can select {\tt take\_training\_program} as the cause and {\tt yr1978\_earning} as the outcome. But if users are interested in the impact of marriage status on income, they can choose {\tt married} as the cause and {\tt yr1978\_earnings} as the outcome. To guide users in reflecting the causal relationships among covariates with the cause and outcome, \confounderdashboard provides a textual explanation of confounders---``{\it a confounder is a third variable that influences both cause and outcome yet does not lie on a causal pathway between cause and outcome}''---when hovering over the question mark next to confounder selection box in Fig.~\ref{fig:system_overview}A. Furthermore, it also provides a table where covariates are ranked by CF scores (ref. Section~\ref{sec:methodology:cf_metrics}) from the highest to the lowest. \tocomments{In the third column, we utilize two side-by-side, vertically aligned histograms \cite{cheng2020dece,ahn2019fairsight} to depict the feature distribution of the untreated group (blue) and the treated group (purple). When $X$ is a continuous variable, it will be discretized into two bins as the ``pseudo'' untreated/treated groups.} 
In \confounderdashboard, users can draw upon multiple information resources, including their domain knowledge regarding the causal structure of cause/outcome and covariates, the statistical CF scores, and the visual information of feature distribution among treatment groups, to select the most likely confounders and put them into the confounder box.

% {After initializing the system and loading data, users can use the \confounderdashboard interface shown in Fig.~\ref{fig:system_overview}A to set up all the required variables for analysis. First, users need to select a pair of cause and outcome variables as their analysis goal. For example, if the user wants to determine if the job training program increases participants' earnings using the aforementioned Lalonde data, they should select {\tt take\_training\_program} as the cause and {\tt yr1978\_earning} as the outcome. If the user is interested in exploring whether marriage status influences participants' earnings, they should set the cause and outcome variables as {\tt married} and {\tt yr1978\_earnings}.}

\subsection{\partition}\label{sec:systemdesign:partition}
The {\partition} interface in Fig.\ref{fig:system_overview}B \tocomments{allows users to create subgroups using two methods}---{\tt MANUAL} and {\tt AUTO}---based on a set of features\footnote{The partitioning variables could also be thought of as confounding variables, but they are special in the sense that they are a subset of the confounders with respect to which we want to capture subgroup heterogeneity.} ({\bf T2}). 
\tocomments{Previous subgroup analysis systems have utilized attribute-value pairs to construct subgroups \cite{cheng2020dece, kwon2022rmexplorer, kwon2020dpvis, guo2023causalvis}. \vispur employs a similar design, enabling users to add or remove covariates and specify the corresponding cut points. As shown in Fig.~\ref{fig:system_overview}B, two variables, {\tt black} and {\tt yr1975\_earning}, are selected, multi-thumb sliders are utilized to determine the cut points. A histogram distribution is displayed above the sliders, providing users with a visual reference for selecting appropriate cut points.}
Alternatively, users can opt for the {\tt AUTO} option, which uses algorithm-supported automated partitioning (ref. Section\ref{sec:methodology:algorithm}). By specifying a few configurations, such as the expected number of subgroups and the minimum size of subgroups, users can easily obtain an algorithm-generated partition that mitigates within-subgroup confounding bias. When clicking the {\tt Submit} button, subgroups based on the partition are generated in \subgroupviewer.

% {${\tt yr1974\_earning \in [7.5, 10.5)}$ $\&$ ${\tt black = True}$, ${\tt yr1974\_earning \in [7.5, 10.5)}$ $\&$ ${\tt black = False}$, ${\tt yr1974\_earning \in [0, 7.5)}$ $\&$ ${\tt black = True}$, as well as ${\tt yr1974\_earning \in [0, 7.5)}$ $\&$ ${\tt black = False}$.}

% DECE: users can refine the group by changing ranges for each feature and click the update button. The users can copy or delete an unwanted subgroup to maintain the subgroup list.
% CausalVIS, RMExplorer: users can click a particular name to facet the visualization by this variable, up to three variables can be selected this way.
% DPVis: a list of pairs of an attribute name and its value range.
\subsection{\subgroupviewer}\label{sec:systemdesign:subgroupviewer}

The {\subgroupviewer} interface in Fig.\ref{fig:system_overview}C provides a comprehensive overview of subgroup patterns, enabling users to understand their heterogeneous characteristics ({\bf T3}). Our design considers subgroup differences not only at the attribute level, but also at the causal behavioral patterns. To achieve this, we have created two views in this panel: {\causalityspace} for exploring causal patterns and {\covariatespace} for analyzing attributes. To ensure a seamless user experience, these two spaces are coordinated with consistent color codings over subgroups. Users can select a subgroup in either space and the interactions will be reflected in both spaces, or hover over subgroups to examine more detailed information.

% {Referring to the causal framework in Fig.\ref{fig:teaser}B (ref. Section~\ref{sec:methodology:diagram}), we focus on subgroup-level {\it propensities}---how likely people in a chosen subgroup are to participant in the training program, {\it base effects}---what are the annual earnings individuals in a subgroup typically obtain in 1978, and cause-outcome {\it associations}---to what extent cause and outcome are correlated.}

\begin{figure}[ht]
    \centering
    \includegraphics[width=0.9\linewidth]{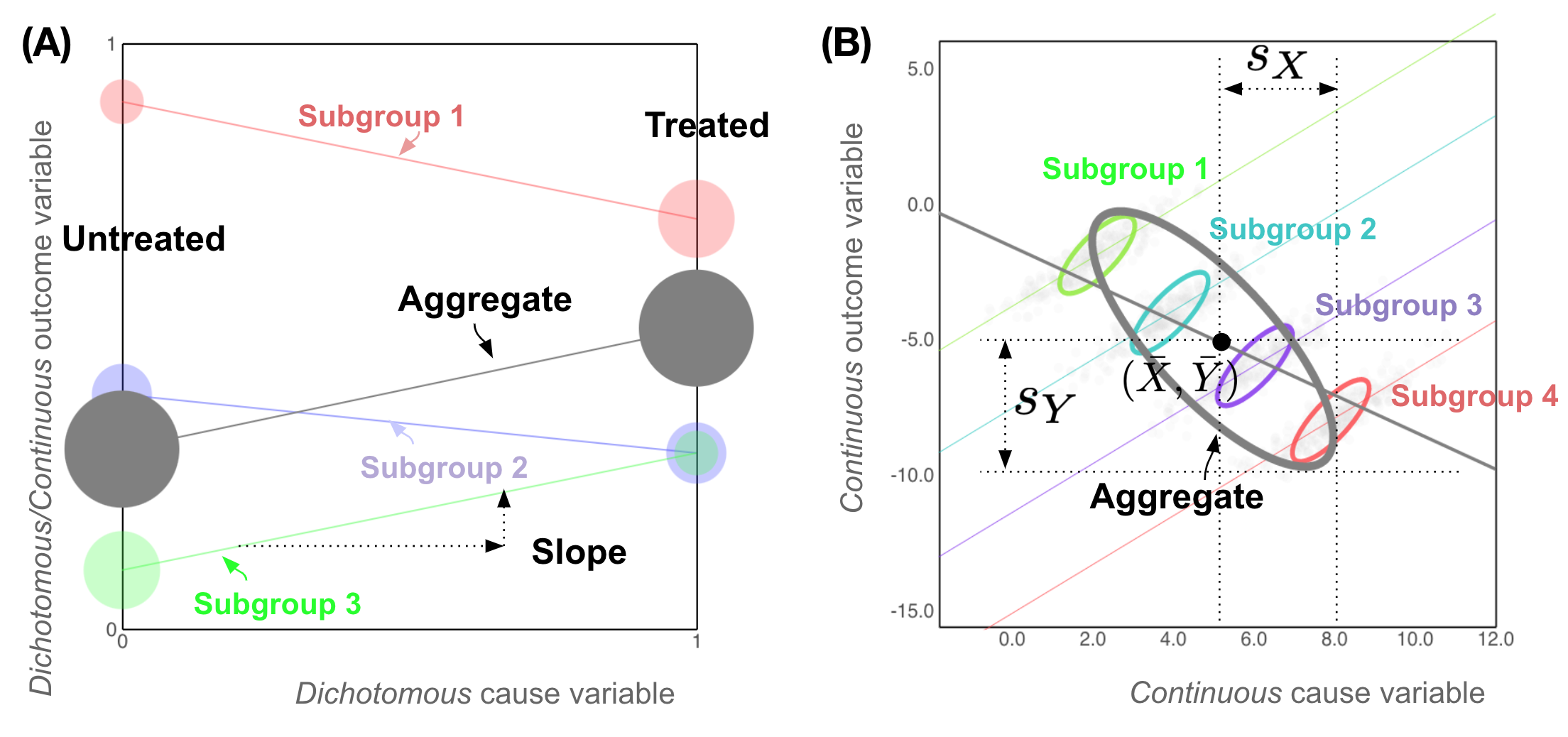}
    \caption{In \subgroupviewer, two visual designs represent the {\causalityspace} for subgroups. The circle-line design (A) is used for dichotomous causes, and the ellipse design (B) is used for continuous causes and outcomes. Both designs use a 2D space with cause on the horizontal axis and outcome on the vertical axis. Subgroups are represented by circle-lines or ellipses with a regression line, and gray elements represent the population while colored elements represent subgroups.}
    \label{fig:2dspace}
\end{figure}

\subsubsection{\causalityspace} This view is shown in Fig.~\ref{fig:system_overview}C, illustrating subgroups' patterns in terms of {\it propensity}, {\it base effect}, and {\it association}. In particular, it represents subgroups in a two-dimensional coordinate space with the horizontal axis as cause and vertical as outcome. Depending on data types, subgroups are represented by either the circle-line design \cite{rucker2008simpson} or the elliptical geometry design \cite{friendly2013elliptical}, as shown in Fig.~\ref{fig:system_overview}C-1, C-2.

{The {\bf circle-line design} (Fig.~\ref{fig:2dspace}A) \tocomments{has been proposed to visualize Simpson's paradox \cite{rucker2008simpson}, particularly depicting cause-outcome} associations where \tocomments{cause is dichotomous and outcome is either dichotomous or continuous.} Each pair of two circles connected by a line represents a subgroup, whose members are divided into control and treated group each represented as a circle with its size indicating the number of subjects. When comparing the sizes of two connected circles, it reveals the treatment propensity (e.g., subgroup 1 has a stronger propensity than subgroup 3). Circles are positioned along two-sided vertical axes indicating the average value of outcome, with the slope between two circles revealing cause-outcome associations. By comparing the distinguished aggregate (grey) against the subgroup-level circle-lines, users can quickly locate a paradoxical phenomenon, such as the aggregate is positive while subgroup 2 is negative.}
{The {\bf elliptical geometry design} \tocomments{is discussed in \cite{friendly2013elliptical} to visualize Simpson's paradox, for cases} where \tocomments{both cause and outcome are continuous} (see Fig.\ref{fig:2dspace}B). The plot consists of colorful ellipses, each representing a subgroup, while a gray ellipse denotes the aggregate to facilitate comparison between the overall data trend and subgroup-level patterns.
The centroid of each ellipse displays the average values of cause and outcome suggesting propensity and base effect, denoted as $(\bar{X},\bar{Y})$ \cite{friendly2013elliptical}. Additionally, the half-widths of the vertical and horizontal projections of each ellipse geometrically reflect  the standard deviations of cause and outcome, represented by $s_X$ and $s_Y$ (Fig.~\ref{fig:2dspace}). The trend of a regression line that passes through the ellipse centroids is indicative of cause-outcome associations.}

% {In terms of interactions, users could mouse over subgroups to examine more detailed information. We note that, in both designs, we utilize slopes to demonstrate cause-outcome associations so that users are able to grasp diverse trends and to locate paradoxical associations. As the true causal effects are always hidden in observational studies, we provide more information in \decisiondiagnosis interface, in Fig.~\ref{fig:system_overview}E, to assist users to conduct spuriousness diagnosis.}

\subsubsection{\covariatespace}

% In information visualization, a glyph refers to a small and compact graphic representation that represents a data point with multidimensional features. 
% Compared with other multidimensional visualization techniques, such as multidimensional scaling (MDS)17, parallel coordinates19, scatterplot matrices, and various advanced designs for reducing clutter in multidimensional data25 or for representing data from heterogeneous dimensions26–30, glyphs transform mul- tidimensional data features to composite visual prop- erties (such as shape, color, and size), producing various ‘‘visual signatures’’ of data points that reveal more complex data patterns and offer a richer descrip- tion about data points.

This view (Fig.\ref{fig:system_overview}C-3) represents the characteristics of subgroups using multivariate radar glyphs \tocomments{due to its compactness and richness. Unlike other multidimensional visualizations (e.g., scatterplot matrices, parallel coordinates \cite{inselberg2009parallel}), glyph visualization encodes complex data features to compact ``visual signatures'' of distinct subgroups \cite{cao2018z}.} The leftmost glyph represents the entire population and the rest of them represent subgroups. In a radar glyph, the axes represent features with dots along the axes indicating the mean values of those features as a summary. To account for the differences in feature scales, we normalize their values into a uniform range of zero to one using $\bar{F}_j = (F_j - \mathrm{min}(F_j)) / (\mathrm{max}(F_j) - \mathrm{min}(F_j)), \forall j$. For categorical features with $J$ discrete values, we convert them into $J - 1$ binary features. In the radar glyphs, we only show the most discriminative top-$k$ features (e.g., $k=5$). To measure a feature's discriminativeness, we trained a series of one-vs-rest binary classifiers with the chosen feature $F$ as input and subgroup labels as output. We then computed an aggregated AUC score as a measurement of feature discriminativeness, where a higher AUC value suggests a stronger ability to distinguish data points from different subgroups. Users have the freedom to select features to investigate by clicking ``Choose Axes.'' E.g., the five axes in Fig.\ref{fig:system_overview}C-3 include {\tt age}, {\tt hispanic}, {\tt no\_degree}, {\tt education\_years}, and {\tt married}.

% {Users could clearly observe the phenomenon of Simpson's paradox by comparing the aggregate trend with any of the subgroup-level trends. If a subgroup were non-identifiable (ref. \ref{sec:methodology:framework:assumption}), our design would clearly highlight it by showing a single circle--either the treated or the untreated---in Fig.~\ref{fig:2dspace}(a). To support interactive data exploration, \vispur allow users to mouse over a subgroup to see the detailed statistics (e.g., estimated effect size and its bootstrap confidence interval), and to highlight (by clicking) two or more subgroups for between-subgroup comparison.}

\begin{figure*}[ht]
    \centering
    \includegraphics[width=0.95\linewidth]{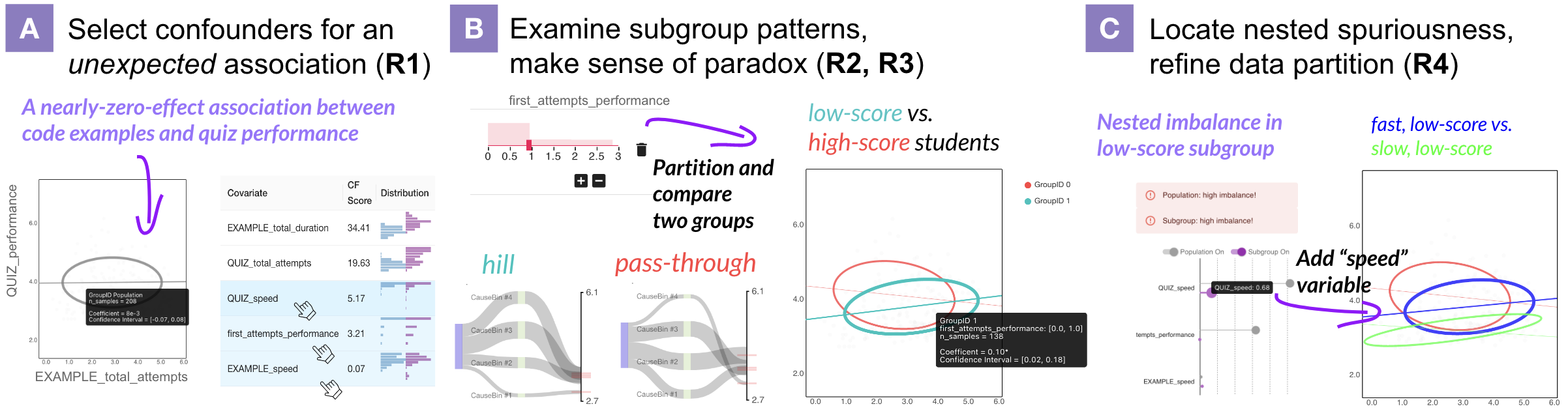}
    \caption{In the case study, Micheal, a education system developer, investigated the effectiveness of using code examples as non-mandatory learning resources in an educational system for Java programming. 
    % {He found a weak association, identified confounders with \confounderdashboard, divided students into subgroups based on prior knowledge using \subgroupviewer, reasoned about initial association using \reasoningstoryboard, detected imbalance with \decisiondiagnosis, and refined partition by adding a nested confounder.}
    % {(A) Initially, he found a weak association and identified prior knowledge and learning patience as potential confounders using \confounderdashboard ({\bf R1}). (B) He divided students into low- and high-score subgroups based on their prior Java knowledge and inspected subgroup patterns using the \subgroupviewer ({\bf R2}). By comparing flow charts in the \reasoningstoryboard, he explained the counterintuitive initial association ({\bf R3}). (C) In the \decisiondiagnosis, he detected residual imbalance and refined his partition by adding a second variable to generate three subgroups for deeper analysis ({\bf R4}).}
    }
    \label{fig:case_study}
\end{figure*}

\subsection{\reasoningstoryboard}\label{sec:systemdesign:reasoningstoryboard}
\begin{figure}[ht]
    \centering
    \includegraphics[width=\linewidth]{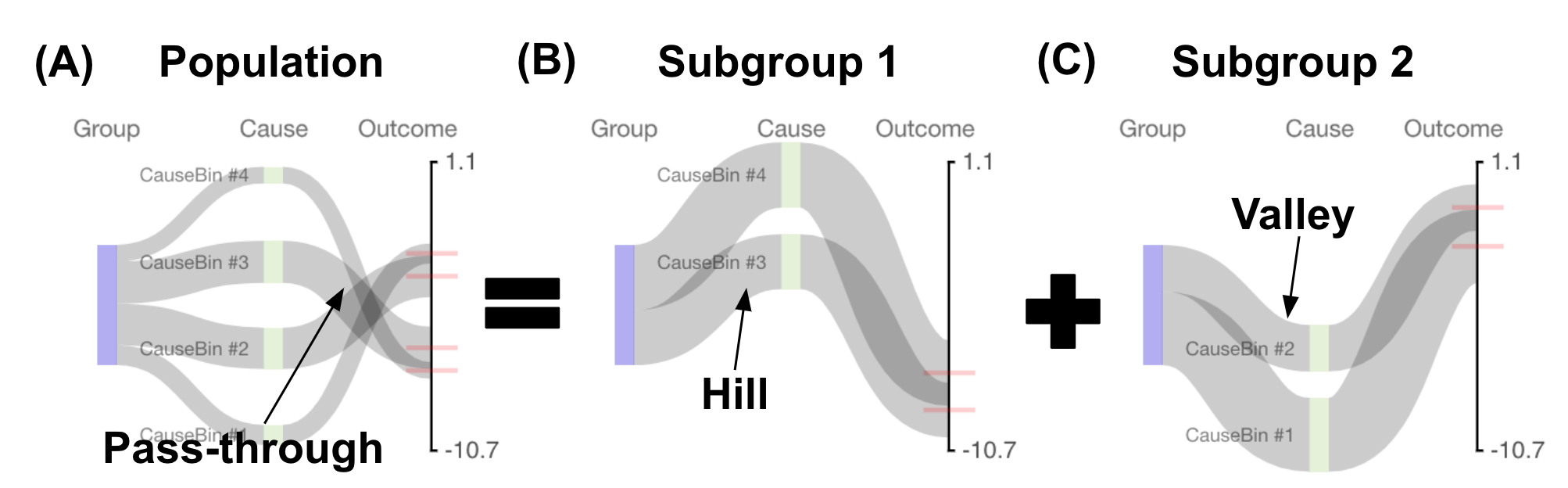}
    \caption{The storyboard design in \reasoningstoryboard follows a flow-based approach. Each diagram includes three layers for {\tt subgroup}, {\tt cause}, and {\tt outcome}. The flow direction reflects the order of participants taking treatment and leading to an outcome, and flow width represents sample size. Three common patterns are displayed: (A) {\bf pass-through}, (B) {\bf hill}, and (C) {\bf valley}.}
    \label{fig:storyboard}
\end{figure}

{To aid in understanding association conflicts, we design {\reasoningstoryboard}, depicted in Fig.~\ref{fig:system_overview}D, which complements {\subgroupviewer} by providing a narrative for the appearance of a conflicting/paradoxical phenomenon ({\bf T4}).} {The diagram comprises three layers: {\tt subgroup}, {\tt cause}, and {\tt outcome}. The {\tt subgroup} node is a rectangle scaled to the size of the chosen subgroup (e.g., the number of participants). In the {\tt cause} layer, multiple nodes depict possible treatments. \tocomments{For a dichotomous treatment, nodes are limited to {treated} and {untreated}; For a continuous treatment, the values are discretized into $L$ ($L = 4$ in our demonstration) bins based on percentiles and ranked in a descending order.} The height of a cause node represents the percentage of participants taking that action. The {\tt outcome} layer is a vertical axis, where the top endpoint indicates the maximum value of the outcome and the bottom indicates the minimum. Pathways originate from the {\tt group} node, traverse through relevant {\tt cause} nodes, and terminate at a specific point along the {\tt outcome} axis. The height of each pathway represents the proportion of participants who took a particular treatment action at each {\tt cause} node. The endpoint of each pathway indicates the average outcome achieved (e.g., average earnings in 1978).}

% little work has studied the flow-based visualization's capability of demonstrating and interpreting a paradoxical phenomenon. Our storyboard design holds contributions by recognizing that the shapes of flows carry valuable information for interpreting paradoxical phenomena.

\tocomments{The storyboard visualization extends the well-known Sankey diagrams \cite{riehmann2005interactive} and parallel sets \cite{kosara2006parallel}, which have been found useful in depicting storylines \cite{kwon2020dpvis} and multidimensional features \cite{kosara2006parallel}. Despite their extensive applications in flow-based visualizations, our contribution lies in presenting paradoxical patterns using flow-based narratives to facilitate the interpretability of such complex phenomena.} Fig.~\ref{fig:storyboard} illustrates three common shape patterns: (A) {\bf pass-through}, (B) {\bf hill}, and (C) {\bf valley}, where (A) depicts the population-level pattern, and (B,C) depict two subgroups' patterns. The {\bf pass-through} shape in Fig.\ref{fig:storyboard}A suggests a {\it negative} cause-outcome association because the subset of subjects taking the largest value of treatment (top) end up having the smallest value of outcome. In contrast, two subgroups in Fig.~\ref{fig:storyboard}B,C exhibit {\bf non-pass-through} (thus {\it positive}) associations. To interpret the reversed associations, users might further investigate Fig.~\ref{fig:storyboard}B,C. Users can see that subgroup (B) tends to take a high-valued treatment but its outcome is relatively low ({\bf hill}), whereas subgroup (C) likes to take a low-valued treatment but its outcome is relatively high ({\bf valley}). When the two mirrored {\bf valley} and {\bf hill} pathways are aggregated together, they could generate a {\bf pass-through} pattern as shown in Fig.~\ref{fig:storyboard}A. {The Lalonde dataset in Fig.~\ref{fig:system_overview}D shows that the valley shape displayed by {\tt GroupID 3} might cancel out the hill patterns of other subgroups, leading to a zero-effect pattern overall. Users can mouse over pathways for detailed information, select subgroup diagrams by clicking the ``add'' icon, and perform shape comparison and paradox reasoning ({\bf T4}).}

\subsection{\decisiondiagnosis}\label{sec:systemdesign:decisiondiagnosis}

To aid decision-making ({\bf T5}), \vispur offers {\decisiondiagnosis} interface (Fig.\ref{fig:system_overview}E) to detect nested confounding bias and spurious associations. Users can select subgroups using the dropdown selector and view statistical results (e.g., effect size, bootstrap CI, $p$-value, sample size) in the {\basicstatistics} table. If the subgroup-level coefficient contradicts the overall association, a warning message---``{Simpson's Paradox}''---is displayed to alert users (Fig.\ref{fig:system_overview}E-1). To detect nested confounding biases in subgroups, we offer the {\imbalancechart} in Fig.~\ref{fig:system_overview}E-2, which computes an {imbalance score} (ref. Section~\ref{sec:methodology:imbalance}) to compare treated and untreated participants within each subgroup. \tocomments{It shares a similar design as the visual components built in Cobalt \cite{greifer2020covariate} and Causalvis \cite{guo2023causalvis} for the purpose of balance checking.} Confounding variables are ranked on the vertical axis by their imbalance score, represented by horizontal position and lollipop size. A warning message will appear when the average imbalance score exceeds a threshold of 0.2, with corresponding lollipops enlarged for visual emphasis. The chart includes two switches for the entire population and the selected subgroup. Users can hover over lollipops to view values and compare pre- and post-partition imbalance scores by toggling the state (on/off) of two switches. {The Lalonde example in Fig.~\ref{fig:system_overview}E-2 demonstrates that partition reduces a large amount of confounding bias caused by {\tt black, yr1974\_earning, yr1975\_earning, married}, however, this chosen subgroup still triggers a warning because of the residual confounder {\tt yr1975\_earning}.}

\tocomments{More details of our iterative development, design choices and considerations are provided in Supplementary Materials \ref{sec:appendix}.}
\section{Case Study \tocomments{In A Real-World Application}: Use and Impact of code examples on students' Java programming skills}\label{sec:casestudy}
\tocomments{In our case study, we worked with {\bf P3}, an educational system designer named Micheal (pseudonym), who specializes in adaptive educational systems and educational data mining. Micheal's objective was to determine whether the Java system developed by his team effectively enhances students' Java programming skills. Fig.~\ref{fig:case_study} demonstrates the process of utilizing \vispur to analyze the relationship between a student's performance and their usage of the system.}

% {We present a case study of an educational system for students to learn Java programming to showcase the effectiveness of \vispur in analyzing how a student's performance is related to use of the system (Fig.~\ref{fig:case_study}). Our interview was with the educational system designer, {\bf P3}, who has expertise in developing adaptive educational systems and educational data mining.}
% {In the given task, he wanted to inspect whether and to what extent, code examples -- provided as non-mandatory learning resources in a developed system -- were beneficial and engaging for students in Java programming education.}

{\bf Settings.} \tocomments{We utilized an educational system dataset provided by Michael, ensuring that its data size and feature dimensions match the real data analysis tasks faced by our target users.} It contains the interactions of 482 undergraduate students with the system who take an introductory course in Java programming from 10 different classrooms during the years of 2020 and 2021. Students used the tool in a non-mandatory manner to study code examples and solve quiz tasks at their own paces, needs, and time. The aim of the data analysis was to determine whether system engagement (measured by the total number of code examples studied by a student {\tt EXAMPLE\_total\_number}) affects a student's performance (measured by the success rate of quiz, {\tt QUIZ\_performance}). The remaining variables include a student's success rate during the first three task attempts, time spent in studying examples and doing quiz, as well as learning speed.

\textbf{Find a counterintuitive association and identify confounders (R1).} He initially observed that the number of code examples studied was not significantly associated with the quiz performance in the \subgroupviewer's {\causalityspace}, which was contrary to his expectation of a positive association. By consulting the table in the \confounderdashboard, he was able to hypothesize potential confounding effects from three variables: {\tt EXAMPLE\_speed}, {\tt QUIZ\_speed}, {\tt first\_attempts\_performance} (Fig.~\ref{fig:case_study}A). Michael explained that {\tt first\_attempts\_performance} might represent a student's prior knowledge of Java, {\tt QUIZ\_speed} and {\tt EXAMPLE\_speed} might indicate a student's learning patience, all of them were potential confounders that might influence a student's attitudes towards the tool and success rate.

{\bf Discover and investigate patterns displayed by different student subgroups (R2).} Micheal decided to use {\tt first\_attempts\_performance} (Fig.~\ref{fig:case_study}B) to divide students into two subgroups: those who failed the initial three tasks (low-score students), and those who answered at least one of them (high-score students). He observed two ellipses with opposite trends in \subgroupviewer, and by hovering over the two ellipses, he confirmed that the positive association for low-score students is significant. He explained that the positive trend in low-score group was very promising, possibly suggesting that the education system was helping those students to improve skills. He also noticed that the ellipse of low-score students was positioned to the lower right of the one of high-score students. He commented: ``{\it Low-score students showed more interest in our code examples, probably because they are not confident and want to learn more.}'' and continued: ``{\it But still, on average, their success rate is lower than that of those capable students.}'' Then he moved to the glyphs in {\covariatespace} to examine how two subgroups differ from each other. He particularly selected two speed-related variables and observed that low-score students clicked faster than high-score students in quiz tasks. He found it reasonable as low-score students tend to quickly click to move on probably because they do not know the answers.

{\bf Reasoning paradoxical associations (R3).} When asked how to explain the paradoxical associations observed in the aggregated data against subgroup data, Micheal moved to \reasoningstoryboard interface to compare two subgroups' flow charts. He observed a hill pattern in low-score chart whereas a valley pattern in high-score chart (Fig.~\ref{fig:case_study}B). Micheal explained: ``{\it Low-score students are more interested in studying code examples. Although they benefit from using the system, they cannot outperform high-score peers eventually, so the aggregated data still shows a negative trend.}''

{\bf Locate nested spuriousness and refine data partition (R4).} Given the mixed effects of two subgroups, Micheal wanted to know more details particularly for the subgroup of low-score students. Through {\imbalancechart} in \decisiondiagnosis, he learned that {\tt QUIZ\_speed} was a nested confounder with a large imbalance score. He added {\tt QUIZ\_speed} as an additional variable and found that low-score students were further divided into slow versus fast learners (Fig.~\ref{fig:case_study}C). Micheal also tried the automated data partition function by setting the number of subgroups to be 3, and a similar data partition was displayed in Fig.~\ref{fig:case_study}C. Micheal explained that, ``{\it high-score students are patient in answering questions in quiz; however, within the low-score subgroup, some students might be more motivated than the others.}''

% {In summary, we demonstrate how \vispur is able to identify distortion and explain the counter-intuitive association, as well as reveal a mixed causal effect pictures over multiple student cohorts. We interviewed a domain expert.}

\section{Controlled User Experiment}\label{sec:userstudy} 
% \toRtwo{Justifications for experiment choice. E.g., why college application example, rationale behind this selection? Dataset size, task difficulty were representative of what the intended users encounter on a daily basis?}

We designed and conducted a controlled user experiment to evaluate \vispur's ability to support a variety of tasks in face of spurious association and Simpson's paradox. We compared \vispur with a contingency table augmented with bar charts as the baseline (see Table S2 in Supplementary Materials~\ref{sec:appendix}), which is a traditional visualization method that has been widely used in describing Simpson's paradox \cite{lesser200111}. We first describe the within-subject study design and then report the results and discuss our quantitative and qualitative findings.

\subsection{Study Design}\label{sec:userstudy:design}
{\bf Participants and procedure.}
We recruited 21 participants, 10 females, 11 males, in our study. Among them, 19 were self-reported to between 25 and 34 years old, 12 hold a Master's degree, and 7 Bachelor's. The participants have a basic knowledge of statistics and are familiar with topics such as correlation, randomized experiments, as well as linear regression. During a 60-minutes of virtual session, we first gave participants a background and system tutorial (20 minutes). We then allowed participants to explore the system and verbally answer four questions for practice (15 minutes).\footnote{{The practices are designed only to ensure that participants fully understand \vispur's interactive features.}}  Afterwards, each participant was given two task scenarios (described below). As we conducted a within-subject study, each participant used \vispur and baseline tool in a random order, and the two task scenarios are randomly shuffled as well. For each task scenario, participants were asked to answer four questions through Qualtrics platform\footnote{\url{https://www.qualtrics.com/}}. At last, we debriefed the participants to learn about their comments or suggestions.

{\bf Tasks and questions.}\label{sec:userstudy:tasks} \tocomments{Existing works have identified Simpson's paradox in a number of application domains, such as college admission \cite{bickel1975sex}, online education \cite{lerman2018computational}, and employment \cite{lalonde1986evaluating,dehejia2002propensity,dehejia1999causal}. Inspired by those examples, we formulate two task scenarios in similar high-stake social settings}: (1) In a high school, a college application training program was developed to help students in getting admitted into colleges. Participants were asked to explore whether the association between ``taking college application training'' and ``being admitted into colleges'' is spurious. (2) An online course was launched to help students to achieve better grades in an exam. Participants were asked to explore whether the association between ``taking online course'' and ``passing the exam'' is spurious. \tocomments{To ensure that the simulated datasets closely resemble what a domain expert might encounter, we constructed two datasets with 6 features and 1500 data records.} In each task scenario, participants needed to answer 4 questions ({\bf Q1---Q4}) with 15 sub-questions in total, designed based on the proposed four design requirements ({\bf R1---R4}):
\begin{enumerate}
    \item[{\bf Q1.}] Identifying confounders ({\bf R1}).
    \item[{\bf Q2.}] Characterizing subgroup difference ({\bf R2}).
    \item[{\bf Q3.}] Interpreting spurious associations or/and Simpson's paradox ({\bf R3}).
    \item[{\bf Q4.}] Making a final decision in face of spuriousness ({\bf R4}).
\end{enumerate}
For example, to test {\bf Q2} we asked: ``{\it how do you agree with the description that on average students from high education family are more likely to be admitted into colleges than students from low education family?''} Questions are listed in Table S3 and S4, Supplementary Materials~\ref{sec:appendix}. In each sub-problem, participants were asked to select the answers in a Likert scale from 1 (strongly disagree or very unlikely), to 5 (strongly agree or very likely). We measured students' performance in terms of both {\bf accuracy} and {\bf certainty} (to what extent participants were confident about their answers when using a tool to search for answers). For our within-subject study, we conducted a paired Wilcoxon signed-rank test (alternative hypothesis is ``one-sided'') to determine whether there was a significant performance margin between participants using \vispur against using baseline.

% {We measured students' performance in terms of both {\bf accuracy} and {\bf certainty} (to what extent participants were confident about their answers when using a tool to search for answers). To measure {accuracy}, we encoded 5 Likert scores into three labels (positive, negative, unknown). Specifically, the scores 1 (strongly disagree or very unlikely) and 2 (somewhat disagree/unlikely) were labeled as negative, 4 (somewhat agree/likely) and 5 (strongly agree or very likely) as positive, and 3 as unknown. To measure {certainty}, we labeled Likert scores 1,5 as certain, and 2,3,4 as uncertain. We measured {\bf accuracy} as the percentage of correct answers among all the 15 sub-questions, and {\bf certainty} as the percentage of certain answers among all the 15 sub-questions. To examine how \vispur, compared to baseline, could improve participants' performance in different types of tasks ({\bf Q1---Q4}), we also computed question-specific accuracy and certainty. For our within-subject study, we conducted a paired Wilcoxon signed-rank test (alternative hypothesis is ``one-sided'') to determine whether there is a significant performance margin between participants using \vispur against using baseline.}

\begin{figure}[ht]
    \centering
    \includegraphics[width=\linewidth]{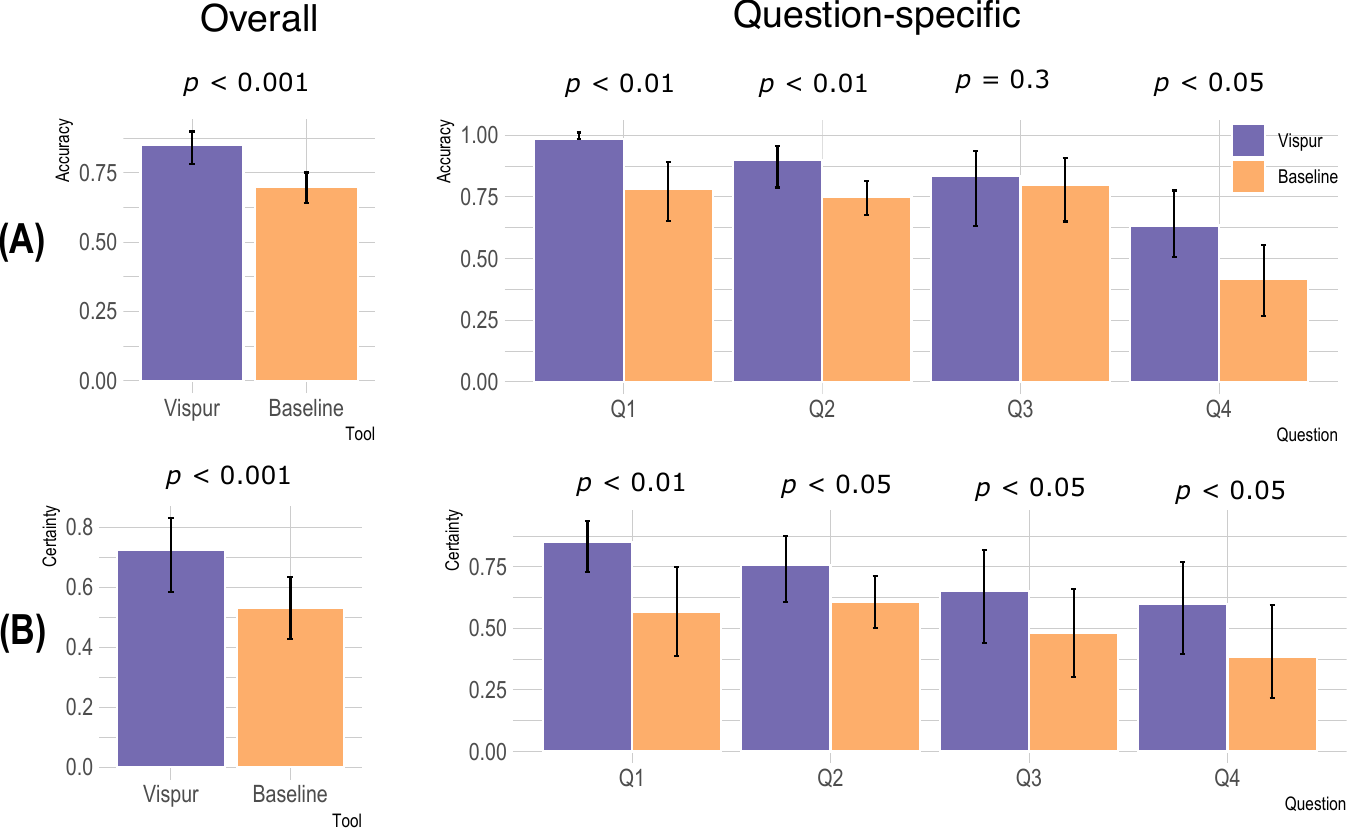}
    \caption{Mean and confidence interval (obtained from R = 999 bootstrap samples) for (A) answering certainty and (B) answerng certainty. Bar chart indicates mean, error bar means CI, $p-$values are obtained from the associated paired Wilcoxon signed-rank test comparing \vispur and baseline.}
    \label{fig:results}
\end{figure}

\begin{figure}[ht]
\centering
\includegraphics[width=\linewidth]{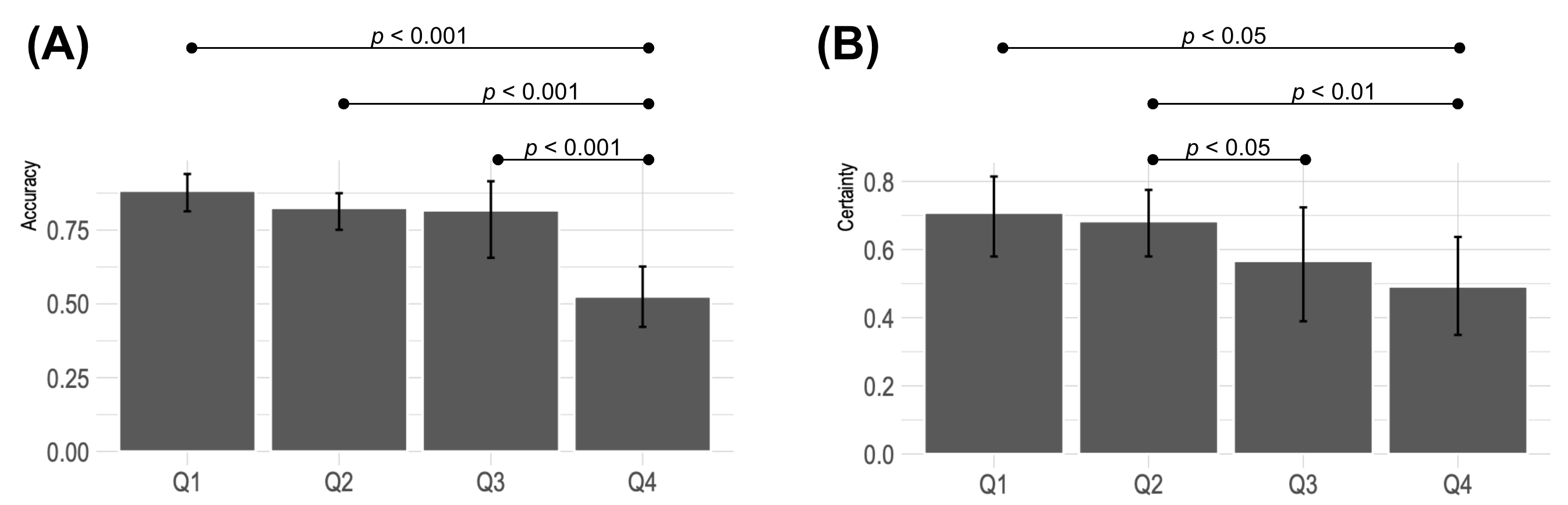}
    \caption{Comparison of participants' performance, reflected by accuracy (A) as well as certainty (B), on four questions {\bf Q1---Q4}. Bar chart indicates mean, error bar means confidence interval. The marked $p-$values are obtained from the corresponding paired Wilcoxon signed-rank test by comparing user performance across any of the 12 question pairs (e.g., {\bf Q1} vs {\bf Q4}).}
    \label{fig:qst_comparison}
\end{figure}

\subsection{Results: Accuracy and Certainty}\label{sec:userstudy:results}
{\bf Result 1. Participants obtained a higher accuracy using \vispur than baseline.} {As shown in Fig.~\ref{fig:results}A, the overall accuracy of \vispur (85.0\%, CI: [78.2\%, 89.0\%]) is significantly higher than that of baseline (70.0\%,  CI: [64.2\%, 75.1\%]) ($V=149.4, p < 0.001$).} {The question-wise accuracy values of \vispur are also higher than that of baseline. Particularly, \vispur has a higher accuracy with a significant margin in {\bf Q1}, {\bf Q2}, and {\bf Q4}.}

{\bf Result 2. Participants were more confident in their answers when using \vispur than using baseline.} {Fig.~\ref{fig:results}B shows that the average certainty score for \vispur is 72.3\% (CI: [58.5\%, 83.3\%]), while for baseline it is 53.0\% (CI: [42.5\%, 62.9\%]). A Wilcoxon signed-rank test suggests a significant difference ($V = 126.5, p < 0.01$). Fig.~\ref{fig:results}B demonstrates a decline in participants' certainty while answering questions from {\bf Q1} to {\bf Q4}, but \vispur are more effective than baseline in boosting participants' certainty over all four questions ($p < 0.05$).}

% {In summary, the evaluation results offer evidence that \vispur, compared to the baseline visualization, was more effective in assisting participants in answering four tasks ({\bf Q1---Q4}) in identifying confounders, characterizing subgroup differences, interpreting spurious associations, as well as making good decisions.}

\subsection{Behavioral Patterns and Qualitative Feedback} We also analyzed the behavioral patterns of participants in using two systems, and gathered qualitative feedback (verbal) from them. \tocomments{Our observations suggest several potential findings that indicate how the design of \vispur might have facilitated} a more intuitive, comprehensive, and interactive exploration of spurious associations, as well as a better understanding of the emergence of paradoxical phenomena.

{\bf Finding 1. \tocomments{\vispur has the potential to reduce cognitive burdens for participants by providing clear and straightforward visualizations.}} 
\tocomments{Participants found \vispur to be a ``{\it useful and simple interface system}'' compared to the baseline, as it reduced their cognitive load. Unlike the baseline tables requiring mental computations, over two-thirds of participants reported faster answer retrieval and increased confidence using \vispur. They appreciated how \vispur presented information simply through positions, directions, and sizes.}
% {Many participants agreed that \vispur is a ``useful and simple interface system'' compared to baseline, as it greatly reduced their cognitive burdens. Based on our observation, nearly 100\% participants have done some mental computations given the raw numbers in baseline tables. In contrast, more than two thirds of participants reported that, in \vispur, they were able to find question answers in a much faster way, and they felt much more confident about their answers, e.g., ``{\it \vispur is easy to use because I had to search for where to look at in baseline tables, but \vispur can directly show me what I need through positions, directions, and sizes.}''}

% {``{\it \vispur helps me find information faster}'', ``{\it \vispur saved time for computing by myself, make me more certain when answering a question by seeing the clear visualizations,}''}

{\bf Finding 2. \tocomments{\vispur has the potential to enhance participants' awareness of} confounding bias and subgroup patterns.} 
Participants reflected on the causal relationships among variables while selecting confounders. They actively compared and analyzed the glyphs, circles, and slopes of different subgroups in the \subgroupviewer and examined the flow charts in the \reasoningstoryboard. A participant commented: ``{\it The baseline visualization is not reliable because hidden information might be overlooked in this table,}'' but ``{\it \vispur provides more functionalities and richer analyses}.''
% {When answering {\bf Q1}, most \vispur users explained how they thought of the causal relationships among variables before clicking to select confounders. For {\bf Q2}, participants actively compared the glyphs, circles, slopes of different subgroups in \subgroupviewer, and inspected the flow charts of those subgroups in \reasoningstoryboard. Through interactions and multiple views, they reported to have obtain a richer understanding with respect to distortion factors and subgroup structure. A participant commented: ``{\it The baseline visualization is not reliable because hidden information might be overlooked in this table,}'' but ``{\it \vispur provides more functionalities and richer analyses}.''}

{\bf Finding 3. \tocomments{Participants found the task of making causal decisions ({\bf Q4}) more challenging compared to {\bf Q1, Q2, Q3}.}}
Some participants expressed difficulty in making a decision for {\bf Q4} and selected ``neither agree nor disagree.'' To further investigate this observation, we computed the aggregated performance scores (accuracy and certainty) for all four questions using both \vispur and the baseline. Fig.~\ref{fig:qst_comparison} reveals that the accuracy of {\bf Q4} is significantly lower than that of other questions {\bf Q1---Q3} ($p < 0.001$), and participants' certainty for {\bf Q4} is also significantly lower than that of {\bf Q1, Q2} ($p < 0.05$).

\section{Discussion}\label{sec:discussion}

% \tocomments{TODO. Revise this section focusing on 3 points: (1) causal knowledge and human beliefs, (2) Limitations in real-world applications, (3) Future work to embed VISPUR as an interactive widget into computational programming environments.}
% {\bf Usability of \vispur's components.} Participants and experts provided positive feedback on \vispur's usability and effectiveness. The \confounderdashboard and \subgroupviewer, along with \partition, were heavily used to examine confounders and subgroup effects. Users found \vispur's flow charts intuitive for explaining Simpson's paradox. The warning messages and imbalance charts in \decisiondiagnosis triggered further thinking and helped identify residual confounders. \vispur was also praised for hypothesis generation and result delivery. Overall, users appreciated the ease of modifying partitions and discussing visualizations.

\tocomments{{\bf Human beliefs and prior causal knowledge.} While \vispur provides qualitative scores and visual signals for confounder identification, it is important to consider human causal knowledge. Relying solely on statistical criteria can lead to the adjustment of undesired variables and introduce bias \cite{hernan2020causal,greenland2003quantifying}. \vispur could incorporate a causality reflection panel \cite{yen2019exploratory} allowing users to draw DAGs before exploring visualizations and data, promoting causal reflection and integration of domain knowledge. Meanwhile, we observed that participants sometimes relied on their prior beliefs in decision-making, overlooking critical visualizations. For example, one participant stated that ``{\it taking a training program shouldn't be bad},'' leading them to always recommend it for individuals in the job market. This confirmation bias \cite{yen2019exploratory} can result in errors as humans tend to ignore information that contradicts their beliefs. One possible solution to counteract individual bias is to encourage collaboration among multiple users. \vispur could support collaboration by allowing people to highlight visualizations, provide interpretations, and engage in discussions with others to collectively reach conclusions.}

\tocomments{{\bf Scalability and applicability in real-world scenarios.} \vispur might encounter two potential limitations in real-world applications. The first limitation is scalability. When users generate a large number of subgroups, the {\subgroupviewer} will become crowed and hard to read. Future work can address this by introducing flexible subgroup operations like hierarchical grouping, filtering, hiding, highlighting, and zooming, to mitigate overlap and information overload. The second limitation pertains to the violation of the causal diagram presented in Figure~\ref{fig:teaser}B. Real-world scenarios often involve complex causal structures, including instrument variables, mediators, and even violations of the four major assumptions (ref. Section~\ref{sec:relatedwork}). A causality reflection panel, as mentioned earlier, could serve as an initial step to assess the alignment between real-world scenarios and our problem setup.}

\tocomments{{\bf Potential of embedding \vispur in programming environments.} Embedding interactive widgets in computational environments (e.g., Jupyter notebooks) has gained popularity among data scientists \cite{kery2020future,wang2022nova}. A future \vispur widget could provide an interactive graphical user interface (GUI) that seamlessly integrates programming and interactive operations. The widget can allow codeless operations that can be synchronized with the notebook, enabling a seamless transition between the visual interface and the coding environment. This integration offers benefits in exploratory data analysis, collaboration, and iterative operations, eliminating the need for users to switch between web-based visual tools and programming environments.}

\section{Conclusion}\label{sec:conclusion}

% {We introduced \vispur, a novel visual analytic system that offers effective methods and informative visual components to aid in the interpretation of spurious associations, with a specific focus on addressing Simpson's paradox. Our system seeks to establish a ``de-paradox'' workflow to combat two primary sources of spuriousness and paradoxical associations: confounding bias and heterogeneous subgroups. By promoting awareness and understanding of spurious associations, \vispur aims to assist data practitioners in making informed data-driven decisions. The evaluation demonstrated the utility and effectiveness of \vispur against existing tools in four key areas: confounder identification, subgroup examination, paradox interpretation, and decision making. Feedback and behavioral patterns indicated that the system is capable of reducing cognitive burdens with clear and intuitive visualizations, allowing users to reason about paradoxical phenomena and practice accountable decision-making. From evaluations, we proved that the novelty of \vispur lies in its ability to provide a comprehensive approach to tackling spurious associations and addressing Simpson's paradox, making it a valuable tool for data practitioners.}

{We present \vispur, a novel visual analytic system that helps interpret spurious associations, focusing on addressing Simpson's paradox. Our system establishes  a ``de-paradox'' workflow, combating confounding bias and heterogeneous subgroups. \vispur enhances awareness and understanding of spurious associations, assisting data practitioners in making informed decisions. The evaluation demonstrates \vispur's utility and effectiveness in confounder identification, subgroup examination, paradox interpretation, and decision making. Feedbacks and user behaviors confirm that \vispur reduces cognitive load with clear visualizations, enables reasoning about paradoxes and accountable decision-making, making it a valuable tool for data practitioners.}

% {In the future, we plan to extend the current design to better address the three limitations discussed above. To better support subgroup investigation when a large number of subgroups are present, we plan to consider interactive operations so that users might freely search, filter, retrieve, highlight and zoom in/out subgroups. We will improve the system's interface to make it more self-contained and user-friendly by inserting self-exploration guidelines. Besides, we plan to develop a more intuitive visual representation for the concept of imbalance to help users identify residual confounding.}

\bibliographystyle{abbrv-doi-hyperref}

\bibliography{main}
\appendix % You can use the `hideappendix` class option to skip everything after \appendix
\section{A List of Supplementary Materials} \label{sec:appendix}
\begin{figure*}
\centering
\includegraphics[width=\linewidth]{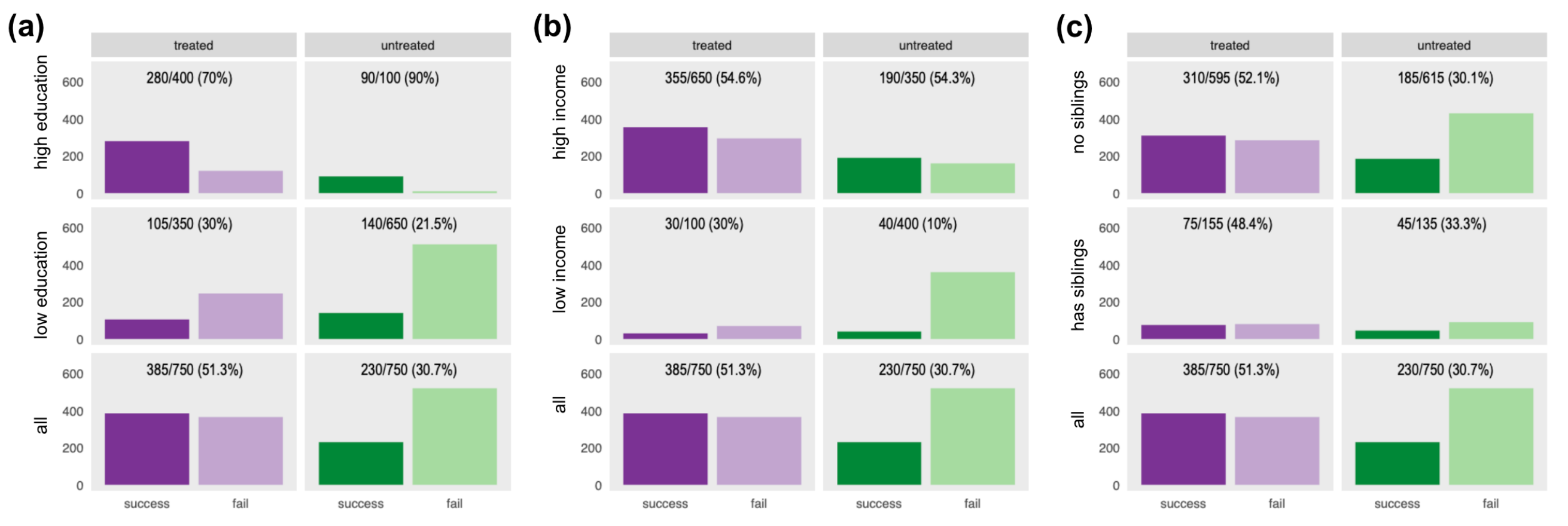}
\caption{The baseline visualization---contingency table augmented by bar charts---used in our user study. It shows data from the first task: whether taking a college training program would increase a student's chance of getting admitted into colleges. Subplots (a-c) are three independent tables where the subjects are divided into subgroups based on parent education level, family income, as well as sibling status. Rows indicate subgroups, and columns are treatment status. The reported raw numbers, such as 280/400 (70\%) in the upper left cell, show that among the total 400 students, 280 out of them were successfully admitted into colleges, thus the success rate is 70\%. Bar charts of two color tones demonstrate the number of students who got a college offer (dark) versus didn't get a college offer (light).}
\label{fig:baseline}
\end{figure*}

Supplementary material are available at \href{https://drive.google.com/drive/folders/1G6PpcE9TOCEdOjc2WWBd0K0Q8Fc6Rin6}{{https://rb.gy/olib8}}. If link does not respond, please copy and paste it into browser. Supplementary materials include:
\begin{itemize}
    \item \tocomments{System design choices and considerations.}
    \item \tocomments{How participants' backgrounds affect their performance in using \vispur.}
    \item The summary of the interview study in Table \ref{tab:interviews} (ref. Section~\ref{sec:designguideline}).
    \item Illustration of the baseline visualization (contingency table augmented by bar charts) used in the controlled user experiments in Fig.~\ref{fig:baseline} (ref. Section~\ref{sec:userstudy}).
    \item Task questions in user study, including (1) Table~\ref{tab:task1}, the task of analyzing whether a college application training program helps a student in getting college offers (ref. Section~\ref{sec:userstudy:tasks}), (2) Table~\ref{tab:task2}, the task of analyzing whether an online course helps students in pass the final examinations (ref. Section~\ref{sec:userstudy:tasks}).
    % \item \vispur system demo. 
\end{itemize}

\section{Acknowledgements}
We thank the anonymous referees for their useful suggestions. The authors would like to acknowledge the support from AFOSR awards and DARPA Habitus program. Any opinions, findings, and conclusions or recommendations expressed in this material do not necessarily reflect the views of the funding sources.

\clearpage

\section{Supplementary Materials}

\tocomments{{\bf System Design Choices and Considerations.} To design our system, the notion of ``subgroup'' is a central concept, as (1) it is through the aggregation of subgroups that a Simpson's paradox emerges, and (2) this concept bears causal implications in terms of heterogeneous causal effects. In our study, the design of \vispur started with defining the core task: {\it visualizing subgroup heterogeneity} to interpret a paradoxical or spurious association, along with the specific design requirements ({\bf R1---R4}) listed in Section~\ref{sec:designguideline:requirement}. Besides, the causal diagram in Fig.~\ref{fig:teaser}B suggests that our visual design should clearly capture a set of six key elements, including three entities, i.e., cause $X$, outcome $Y$, covariates $Z$ (or group membership), and three arrows, i.e., propensity $\mathbf{Z}\rightarrow T$, base effect $\mathbf{Z}\rightarrow Y$, cause-outcome relationship $X\leftrightarrow Y$. Given such considerations, we investigated five designs in existing works of visualizing Simpson's paradox, including the B-K diagram \cite{baker2001good}, platform scale \cite{rum1980magic}, comet \cite{armstrong2014visualizing}, circle-line design \cite{rucker2008simpson}, as well as ellipse design \cite{friendly2013elliptical}. Table~\ref{tab:comp} provides a summary of advantages and disadvantages of the five visual diagrams examined in our design decision. Table~\ref{tab:visulizing_SP} summarizes whether and how the six key elements are represented in those five visual designs.}

\tocomments{Among them, the design of platform scale includes a set of blocks arranged in stacks of varying heights being placed on a platform and balanced on a pivot at the center of gravity \cite{rum1980magic}. Although the gravity-based design is intuitive, it is difficult to visually comprehend how a static equilibrium could be achieved when more than two subgroups (i.e., stacks of blocks) are present. As the blocks of the same subgroup are placed on different platforms (corresponding to different treatment options), it is not easy to examine the properties of a chosen subgroup. All of the other four designs have utilized a two-dimensional coordinate space, where the vertical axis represents outcome, but the horizontal axis holds different meanings in different designs. For example, the B-K diagram considers the horizontal axis as the percentage of a chosen subgroup (two subgroups in total), therefore it does not have a explicitly visual representation of more than two subgroups, failing to satisfy our design requirement \cite{baker2001good}. The design of comet consists of a set of comets (i.e., subgroups) placed in a two-dimensional space, and the motion of a comet from head to tail captures its change from the starting time point A to the ending time point B \cite{armstrong2014visualizing}. Although it well preserves the notion of subgroup and enables a straightforward comparison of subgroup patterns, the strong sense of motion conveyed by comets only works well for time-based data rather than a more general treatment scenario. We compare the pros and cons of all possible visual representation candidates, and decided that circle-line \cite{rucker2008simpson} and ellipse diagram \cite{friendly2013elliptical} are most suitable to meet the required elements and data types. The first reason is representation consistency. They share the same two-dimensional cause-outcome space, with the horizontal axis as cause and vertical as outcome, and the slope of regression lines as cause-outcome relationship. Being represented in the cause-outcome space, circle-line plot is designed for a dichotomous treatment while ellipse diagram for a continuous treatment. Secondly, the diagram, either a circle-line plot or a data ellipse, is a simple and effective summary of raw data points in the two-dimensional space. They are good at simplifying a possibly large-scale data set and capturing the most important information (e.g., the set of six key elements, including three entities and three arrows, are well represented, see Table~\ref{tab:visulizing_SP}). The third reason is that they enable a good representation and comparison of multiple subgroups. Both provide a high-level overview of multiple subgroups and enable inter-subgroup comparison; besides, an aggregate plot for the entire data (circle-line or ellipse) could be shown to alert Simpson's paradox.}

\tocomments{During the system design and development process, we conducted two rounds of open-ended discussion and feedback collection sessions with 5 colleagues in our research lab. We invited them to raise any concerns and suggestions to improve our system. For instance, the feature histograms in {\confounderdashboard} were initially designed to be overlapped, but users might have difficulty in reading distinctions when two distributions look similar. We decided to put them side by side vertically to give a straightforward comparison. The initial design of multi-thumb sliders didn't contain the associated feature distribution. Users reported that they sometimes had no idea where to put the cutting points. We then follow their suggestions to visualize a histogram distribution to guide the selection of proper cut points. In addition, to design the {\reasoningstoryboard}, we have tested two options to visualize data volume, through flow width versus through color darkness. Our users reported that width is a much clearer visual signal for them to comprehend.}

\begin{table*}
 \centering
 \small
 \caption{\tocomments{Comparison of existing visualization designs for Simpson's paradox. It provides a summary of design ideas, advantages and disadvantages of each of the designs.}}
 \label{tab:comp}
 % \begin{adjustbox}
 \begin{tabular}{ p{0.1 \linewidth} p{0.35 \linewidth} p{0.2 \linewidth} p{0.25 \linewidth} }
    \toprule
    {\bf Graph} & {\bf Design summary} & {\bf Advantages} & {\bf Disadvantages} \\
    \midrule
    {\bf B-K diagram} \cite{baker2001good} & 
    A 2D coordinate space where the horizontal line is the proportion of a certain group (e.g., women) and the vertical line is outcome. Lines in the space demonstrate different treatment, vertical positions indicate outcomes. & 
    Well preserve the outcome differences of distinct treatments using vertical positions of dots or lines. & 
    No clear representation of subgroups, thus does not support comparison of subgroups' properties. \\ \\
    {\bf Platform scale} \cite{rum1980magic} & 
    A set of blocks arranged in stacks of varying heights is located on a platform and balanced on a pivot at the center of gravity. Treatments are represented by multiple platforms, outcomes are by positions of blocks, treatment preferences are by heights of blocks. & 
    Well represent weights (i.e., treatment preferences) by heights of blocks. &
    Hard to visually understand how a static equilibrium is achieved when multiple subgroups (stacks of blocks) are present; Hard to examine subgroup-specific properties as blocks of the same subgroup are placed on different platforms. \\ \\
    {\bf Comet} \cite{armstrong2014visualizing} &
    A set of comets placed in a 2D coordinate space, where each comet is a subgroup, with the thin head being one treatment scenario (A) and the thick tail the other (B). The motion of a comet indicates how the changes of properties from scenario A to B. & 
    Well represent multiple subgroups, and easily compare difference of subgroup patterns, an aggregate comet is shown to alert Simpson's paradox. & 
    The strong sense of motion conveyed by comets only works well for time-based data rather than continuous or categorical treatment. \\ \\
    {\bf Circle-line plot} \cite{rucker2008simpson} &
    A set of circle-line plots placed in a 2D coordinate space, where each circle-line plot is a subgroup with its slope representing per-subgroup association, horizontal axis being treatment, vertical axis being outcome, and circle size represents sample size. &
    Well represent multiple subgroups, and easily compare difference of subgroup patterns, an aggregate circle-line is shown to alert Simpson's paradox. & 
    Large circles obscure small ones if many subgroups exist.  \\ \\
    {\bf Ellipse} \cite{friendly2013elliptical} &
    A set of ellipses placed in a 2D coordinate space, where each ellipse is a subgroup with its slope representing per-subgroup association, horizontal axis being treatment, vertical axis being outcome. &
    Well represent multiple subgroups, and easily compare difference of subgroup patterns, an aggregate ellipse is shown to alert Simpson's paradox. &
    Large ellipses obscure small ones if many subgroups exist. \\
    \bottomrule
 \end{tabular}
 % \end{adjustbox}
\end{table*}

\begin{table*}
\small
\centering
\caption{\tocomments{Comparison of existing designs for Simpson's paradox. It summarizes how six key elements (see Fig.~\ref{fig:teaser}B)---three entities (cause, outcome, subgroup) and three arrows (treatment propensity, base effect, cause-outcome relationship)---are represented using visual encodings in each of the designs.}}
\label{tab:visulizing_SP}
% \begin{adjustbox}{width=\textwidth}
\begin{tabular}{ p{0.1\linewidth} p{0.1\linewidth} p{0.1\linewidth} p{0.1\linewidth} p{0.1\linewidth} p{0.1\linewidth} p{0.1\linewidth}} 
 \toprule
 {\bf Diagram} &  {\bf Subgroup} & {\bf Cause} & {\bf Outcome} & {\bf Treatment propensity} & {\bf Base effect} & {\bf Cause-outcome association} \\
 \midrule
 {\bf B-K diagram} \cite{baker2001good} & - & line segments & y-axis & x-coordinate of data points along line segments & endpoints of line segments & slope of line segments \\ \\
 {\bf Platform scale} \cite{rum1980magic} & block labels & platforms & positions of blocks on platforms & height of blocks & a default stack of blocks & relative positions of blocks with the same label \\ \\
 {\bf Comet} \cite{armstrong2014visualizing} & comets & endpoints of a comet & y-axis & x-coordinate of a comet & y-coordinate of a comet's head & a comet's length along y-axis \\ \\
 {\bf Circle-line plot} \cite{rucker2008simpson} & line segments & 0/1 on x-axis & y-axis & circle size & intercept on y-axis of line segments & slope of line segments \\ \\
 {\bf Ellipse} \cite{friendly2013elliptical} & ellipses & x-axis & y-axis & positions of ellipses along x-axis & positions of ellipses along y-aixs & slope of regression lines \\
 \bottomrule
\end{tabular}
% \end{adjustbox}
\end{table*}

\begin{figure}
\centering
\includegraphics[width=\linewidth]{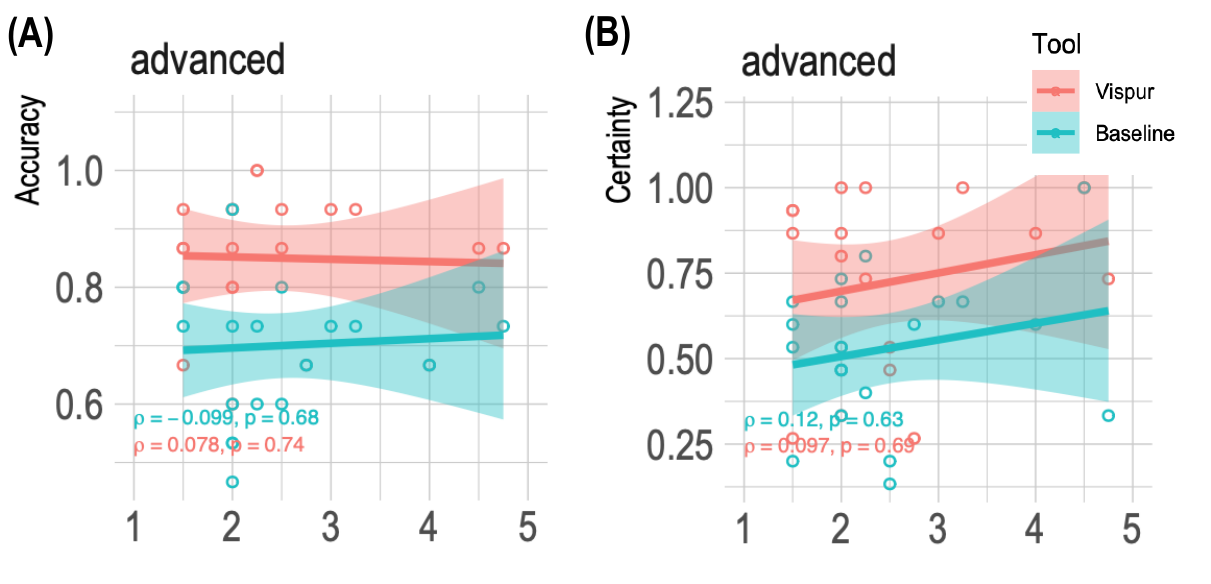}
\caption{The relationship between participants' performance and their self-reported familiarity with advanced causality topics. The x-axis indicates the mean score of participants' self-reported familiarity with advanced causality topics, and the y-axis shows accuracy (A) and certainty (B). The plots show regression lines with a 95\% confidence interval as well as the Spearman's rank correlation coefficient ($\rho$) and the corresponding $p-$values.}
\label{fig:bck_results}
\end{figure}

\tocomments{{\bf How Participants' Backgrounds Affect Performance in Using \vispur.} In the controlled experiments, we asked participants to report their level (1 to 5) of being familiar with a series of advanced causality topics including random experiments, causal inference, confounding, propensity score, and selection bias. We computed the average value of self-reported scores regarding those advanced topics, and computed the Spearman's rank correlation coefficient ($\rho$) between this score and participants' performance. As shown in Fig.~\ref{fig:bck_results}A, the correlation coefficients between expertise score and accuracy for both methods, \vispur and baseline, are not significant ($\rho_{\mathrm{VISPUR}} = 0.078, p = 0.74$, and $\rho_{\mathrm{Baseline}} = -0.099, p = 0.68$). Similarly, we didn't observe significant positive correlations between expertise score and certainty in Fig.~\ref{fig:bck_results}B ($\rho_{\mathrm{VISPUR}} = 0.097, p = 0.69$, and $\rho_{\mathrm{Baseline}} = 0.12, p = 0.63$). To summarize, we didn't find a strong evidence showing that participants of various causality backgrounds have very different performances. We note that our experiments are not free of limitations. For instance, our system is designed for a broader range of data practitioners, so the majority of our recruited participants are not experts of causality. The scatter plots in Fig.~\ref{fig:bck_results} shows that most participants reported to have a familiarity level at 1-3 with advanced causality topics. Recruiting and testing individuals with greater expertise in the field is critical for gaining a more complete understanding of how expert users perform when utilizing VISPUR compared to ordinary data practitioners. Due to the excessive time needed for collecting data from each participant, we did not administer a pre-study knowledge test to participants; therefore, our analyses may contain self-reporting errors.}

\begin{table*}[t]
  \centering
  \small
  \caption{Summary of our interview study. Colors represent four major challenges: \conebox{\bf C1}, \ctwobox{\bf C2}, \cthreebox{\bf C3}, \cfourbox{\bf C4}.}
  \label{tab:interviews}
   % \begin{adjustbox}{width=\textwidth}
   \begin{tabular}{ p{0.10\linewidth} p{0.20\linewidth} p{0.24\linewidth} p{0.36\linewidth}} 
    \toprule
    {\bf Interviewee} & {\bf Focal Interest in Analysis} & {\bf Practices/Tools} & {\bf Challenges/Needs} \\
    \midrule
    {Social worker ({\bf P1})} & {To what extent crime severity is linked to sentence, any differences among different populations (e.g., ethnicity, gender)?} & Fit one or multiple regression models by adjusting demographics features and examine coefficients and $p-$values; SPSS and R are used; & 
    \cone{Have difficulty in variable selection in regression and might overlook confounders;} \cfour{Unable to claim causal effects from statistical analysis;} \ctwo{Unable to automatically detect causal effects for subpopulations especially when multiple variables are involved;} Desire for a user-friendly, interactive system for data exploration; \\
    \midrule
    {Trading analyst ({\bf P2})} & {How and under what special conditions does previous return inform future return in the trading market?} & Build regression models and examine the coefficients and $p-$value; Rely on prior experience to manually search for a subset of market conditions and re-fit regressions; R and Python are used; & 
    \cone{Lack guidance in feature selection;} \cthree{Unable to interpret association change in different models;} \cfour{Unable to claim significance or make decisions given inconsistent associations in different models;} \ctwo{Desire for guidance in searching for market conditions where past return predicts future;} Desire for intuitive visualization to deliver/explain results to leadership; \\
    \midrule
    {Educational system designer ({\bf P3})} & {What is the impact of the designed educational system on students' performance, and how can we explain the counterintuitive pattern of ``more engagement with the system $\rightarrow$ worse performance''?} & Build (stepwise) regression models to examine coefficients and $p-$value; Multicollinearity examination and data-driven feature selection; R and Python are used; & 
    \cthree{Unable to interpret association changes in different models of distinct predictors;} \ctwo{How to define subgroups remain a challenge in face of many features, to reveal students' heterogeneous characteristics (e.g., who like to use the system, how the system affects students' performance);} \cthree{Desire for tools/techniques to explain the counterintuitive association found in their work;} \cfour{Hard to claim a strong conclusion regarding the effectiveness of the designed system;} Interpret the analysis results to people without background of causal inference is a challenge; \\
  \bottomrule
\end{tabular}
% \end{adjustbox}
\end{table*}

\begin{table*}[ht]
  \centering
  \small
  \caption{The task of analyzing whether a college application training program helps a student in getting offers.}
  \label{tab:task1}
   \begin{tabular}{ p{0.95\linewidth} }
    \toprule
    {\bf Task 1. A high school has launched a college application training program, aiming to help students in getting admitted into colleges. You will be using a visual tool to explore a data set and figure out whether the association between ``taking college application training'' and ``being admitted into colleges'' is spurious. Please note, the treated are people who have taken the training program, while untreated those who didn't take it.} \\ 
    {\begin{itemize}
    \setlength\itemsep{1em}
        \item[{\bf Q1}] {\bf [Confounders]} Based on the given data and visualization, how do you agree with the following descriptions about comparing the treated/untreated?
        \begin{itemize}
        \setlength\itemsep{0em}
            \item[(a)] {\it The treated and untreated differ much in their parents' education attribute.}
            \item[(b)] {\it The treated and untreated differ much in their family income attribute.}
            \item[(c)] {\it The treated and untreated differ much in their attribute of ``having sibling.''}
        \end{itemize}
        \item[{\bf Q2}] {\bf [Subgroups]} Suppose the data is divided into two subgroups based on parent education level (i.e., high vs low), how do you agree with the following descriptions?
        \begin{itemize}
        \setlength\itemsep{0em}
            \item[(a)] {\it In the high-education subgroup, more than 50\% individuals have taken the training.}
            \item[(b)] {\it On average students from high education family are more likely to be admitted into colleges than students from low education family.}
            \item[(c)] {\it Focusing on the high education subgroup, the success rate for the treated is significantly lower than that for the untreated.}
            \item[(d)] {\it Combining high-/low-education subgroups together, the success rate for the treated is significantly higher than that of the untreated.}
            \item[(e)] {\it Splitting participants based on parent education, the proportion of students having siblings in the high-education subgroup is significantly higher than that in the low-education subgroup.}
            \item[(f)] {\it The data exhibits Simpson's paradox, as the subgroup-level treatment-outcome association in low-/high-education subgroups are different or even reversed from the overall.}
        \end{itemize}
        \item[{\bf Q3}] {\bf [Reasoning]} Given the observation that the treatment-outcome association is negative in the high-educated subgroup (e.g., a higher success rate among the untreated instead of the treated), which is opposite from the population-level association, how much do you agree with the following explanations for this phenomenon?
        \begin{itemize}
        \setlength\itemsep{0em}
            \item[(a)] {\it The students with high education parents are more likely to take the training while the reverse is true for those with low education parents; meanwhile, they also tend to have a higher likelihood of getting into colleges than the low-educated. These two factors -- ``those who likely to take treatment happen to be those who likely to succeed'' -- result in a spurious positive association overall.}
            \item[(b)] {\it The students with low education parents are more likely to take the training while the reverse is true for those with high education parents; meanwhile, they also tend to have a higher likelihood of getting into colleges than the high-educated. These two factors -- ``those who likely to take treatment happen to be those who likely to succeed'' -- result in a spurious positive association overall.}
            \item[(c)] {\it The students with high education parents are more likely to take the training while the reverse is true for those with low education parents; meanwhile, they also tend to have a higher likelihood of getting into colleges than the low- educated. These two factors -- ``those who likely to take treatment happen to be those who likely to fail'' -- result in a spurious positive association overall.}
        \end{itemize}
        \item[{\bf Q4}] {\bf [Decisions]} Assuming no other hidden confounders involved, or any other mechanisms distorting the cause- outcome relationships, you’re a decision-maker, based on this given data, which decisions you might agree with?
        \begin{itemize}
        \setlength\itemsep{0em}
            \item[(a)] {\it Might recommend a student who has siblings to take the training program. Because the association is positive among the participants having siblings, implying that taking the program can increase the chance of such a student being admitted into colleges.}
            \item[(b)] {\it Might recommend a student who does not have siblings to take the training. Because the association is positive among the participants who do not have siblings, besides, other covariates are balanced between treated/untreated, implying that taking the program would increase the chance of such a student in being admitted into colleges.}
            \item[(c)] {\it Might not recommend a student from high education family to take the training program. Because the association is negative among the participants who are from high education family, besides, other covariates are balanced between treated/untreated, implying that taking the program would decrease the chance of such a student in getting admitted into colleges.}
        \end{itemize}
    \end{itemize}} \\
    \bottomrule
\end{tabular}
\end{table*}

\begin{table*}[ht]
  \centering
  \small
  \caption{The task of analyzing whether an online course helps students in pass the final examinations.}
  \label{tab:task2}
   \begin{tabular}{ p{0.95 \linewidth} }
    \toprule
    {\bf Task 2. An online course is developed, aiming to help students to achieve better grades to pass an exam. You will be using a visual tool to explore a data set and figure out whether the association between ``taking online course'' and ``passing the exam'' is spurious. Please note, the treated are people who have taken the online course, the untreated otherwise.} \\
    {\begin{itemize}
    \setlength\itemsep{1em}
        \item[{\bf Q1}] {\bf [Confounders]} Based on the given data and visualization, how do you agree with the following descriptions about comparing the treated/untreated?
        \begin{itemize}
        \setlength\itemsep{0em}
            \item[(a)] {\it The treated and untreated differ much in their age attribute.}
            \item[(b)] {\it The treated and untreated differ much in their skill attribute.}
            \item[(c)] {\it The treated and untreated differ much in their gender attribute.}
        \end{itemize}
        \item[{\bf Q2}] {\bf [Subgroups]} Suppose the data is divided into two subgroups based on age (i.e., old vs young), how do you agree with the following descriptions?
        \begin{itemize}
        \setlength\itemsep{0em}
            \item[(a)] {\it In the old subgroup, more than 50\% individuals have taken this online course.}
            \item[(b)] {\it On average, the old people are more likely to pass the exam than the young.}
            \item[(c)] {\it Focusing on the old subgroup, the success rate of passing the exam for the treated is significantly lower than that of the untreated.}
            \item[(d)] {\it Combining the old/young subgroups together, the success rate of passing the exam for the treated is significantly higher than that of the untreated.}
            \item[(e)] {\it Splitting the participants based on age, the proportion of female students in the old subgroup is significantly higher than that in the young subgroup.}
            \item[(f)] {\it The data exhibits Simpson's paradox, as the subgroup-level treatment-outcome association in old/young subgroups are different or even reversed from the overall.}
        \end{itemize}
        \item[{\bf Q3}] {\bf [Reasoning]} Given the observation that the treatment-outcome association is negative in the old subgroup (e.g., a higher success rate among the untreated instead of the treated), which is opposite from the population-level association, how much do you agree with the following explanations for this paradoxical phenomenon?
        \begin{itemize}
        \setlength\itemsep{0em}
            \item[(a)] {\it The old participants are more likely to take the online course while the reverse is true for the young; meanwhile, they also tend to have a higher likelihood of passing the exam than the young. These two factors -- ``those who likely to take the treatment happen to be those who likely to succeed'' -- result in a spurious positive association overall.}
            \item[(b)] {\it The young participants are more likely to take the online course while the reverse is true for the old; meanwhile, they also tend to have a higher likelihood of passing the exam than the old. These two factors -- ``those who likely to take the treatment happen to be those who likely to succeed'' -- result in a spurious positive association overall.}
            \item[(c)] {\it The old participants are more likely to take the online course while the reverse is true for the young; meanwhile, they also tend to have a higher likelihood of passing the exam than the young. These two factors -- ``those who likely to take the treatment happen to be those who likely to fail'' -- result in a spurious positive association overall.}
        \end{itemize}
        \item[{\bf Q4}] {\bf [Decisions]} Assuming no other hidden confounders involved, or any other mechanisms distorting the cause- outcome relationships, you’re a decision-maker, based on this given data, which decisions you might agree with?
        \begin{itemize}
        \setlength\itemsep{0em}
            \item[(a)] {\it Might recommend a male student to take the online course. Because the association is positive among male participants, implying that taking the program can increase the chance of a male student getting a job.}
            \item[(b)] {\it Might recommend a female person to take the online course. Because the association is positive among female participants, besides, other covariates are balanced among treated/untreated, implying that taking the program would increase the chance of a female participant in passing the exam.}
            \item[(c)] {\it Might not recommend an old person to take the online course. Because the association is negative, besides, other covariates are balanced among treated/untreated, implying that taking the program would decrease the chance of an old person in passing the exam.}
        \end{itemize}
    \end{itemize}} \\
\bottomrule
\end{tabular}
\end{table*}

% \tocomments{{\bf System Demo.} We provide two demos:
% \begin{enumerate}
%     \item A longer version of VISPUR demo with a brief background (7 min), please click \href{https://drive.google.com/drive/folders/1mBIHysciAV8kvripizKeE2_2sr2deeeG}{https://shorturl.at/fwJT4}.
%     \item A shorter version of VISPUR demo without background information (5 min), please click \href{https://drive.google.com/drive/folders/1mBIHysciAV8kvripizKeE2_2sr2deeeG}{https://shorturl.at/lmqzE}.
% \end{enumerate}}

\end{document}